\documentclass{ieeeaccess}
\usepackage{cite}
\usepackage{amsmath,amssymb,amsfonts}
\usepackage{algorithmic}
\usepackage{graphicx}
\usepackage{textcomp}
\usepackage{bm}
\usepackage{subcaption}
\def\BibTeX{{\rm B\kern-.05em{\sc i\kern-.025em b}\kern-.08em
    T\kern-.1667em\lower.7ex\hbox{E}\kern-.125emX}}

\usepackage{booktabs} 
\usepackage{algorithm,algorithmic}

\usepackage[hyphens]{url}

\usepackage{tikz}
\usepackage{pgfplots}
\usepackage{tikzscale}
\usepackage{tikz-qtree}
\pgfplotsset{compat=newest}
\pgfplotsset{plot coordinates/math parser=false}
\pgfplotsset{every axis/.append style={
                    label style={font=\scriptsize},
                    tick label style={font=\scriptsize},
                    legend style={font=\scriptsize}
                    }}
\usetikzlibrary{plotmarks,shapes,patterns,decorations.pathreplacing,backgrounds,calc,arrows,arrows.meta,spy,matrix,shadows,trees,positioning,fit}
\usepgfplotslibrary{patchplots,groupplots}

\tikzstyle{startstop} = [rectangle, rounded corners, minimum width=2cm, minimum height=0.5cm,text centered, draw=black]
\tikzstyle{io} = [trapezium, trapezium left angle=70, trapezium right angle=110, minimum width=3cm, minimum height=1cm, text centered, draw=black]
\tikzstyle{process} = [rectangle, minimum width=2cm, minimum height=0.5cm, text centered, draw=black, alignb=center]
\tikzstyle{decision} = [ellipse, minimum width=2cm, minimum height=1cm, text centered, draw=black]
\tikzstyle{arrow} = [thick,<->,>=stealth]
\tikzstyle{line} = [thick,>=stealth]
\tikzstyle{darrow} = [thick,<->,>=stealth,dashed]
\tikzstyle{sarrow} = [thick,->,>=stealth]
\tikzstyle{larrow} = [line width=0.1mm,dashdotted,->,>=stealth]

\makeatletter
\def\grd@save@target#1{%
  \def\grd@target{#1}}
\def\grd@save@start#1{%
  \def\grd@start{#1}}
\tikzset{
  grid with coordinates/.style={
    to path={%
      \pgfextra{%
        \edef\grd@@target{(\tikztotarget)}%
        \tikz@scan@one@point\grd@save@target\grd@@target\relax
        \edef\grd@@start{(\tikztostart)}%
        \tikz@scan@one@point\grd@save@start\grd@@start\relax
        \draw[minor help lines] (\tikztostart) grid (\tikztotarget);
        \draw[major help lines] (\tikztostart) grid (\tikztotarget);
        \grd@start
        \pgfmathsetmacro{\grd@xa}{\the\pgf@x/1cm}
        \pgfmathsetmacro{\grd@ya}{\the\pgf@y/1cm}
        \grd@target
        \pgfmathsetmacro{\grd@xb}{\the\pgf@x/1cm}
        \pgfmathsetmacro{\grd@yb}{\the\pgf@y/1cm}
        \pgfmathsetmacro{\grd@xc}{\grd@xa + \pgfkeysvalueof{/tikz/grid with coordinates/major step x}}
        \pgfmathsetmacro{\grd@yc}{\grd@ya + \pgfkeysvalueof{/tikz/grid with coordinates/major step y}}
        \foreach \x in {\grd@xa,\grd@xc,...,\grd@xb}
        \node[anchor=north] at (\x,\grd@ya) {\pgfmathprintnumber{\x}};
        \foreach \y in {\grd@ya,\grd@yc,...,\grd@yb}
        \node[anchor=east] at (\grd@xa,\y) {\pgfmathprintnumber{\y}};
      }
    }
  },
  minor help lines/.style={
    help lines,
    gray,
    line cap =round,
    xstep=\pgfkeysvalueof{/tikz/grid with coordinates/minor step x},
    ystep=\pgfkeysvalueof{/tikz/grid with coordinates/minor step y}
  },
  major help lines/.style={
    help lines,
    line cap =round,
    line width=\pgfkeysvalueof{/tikz/grid with coordinates/major line width},
    xstep=\pgfkeysvalueof{/tikz/grid with coordinates/major step x},
    ystep=\pgfkeysvalueof{/tikz/grid with coordinates/major step y}
  },
  grid with coordinates/.cd,
  minor step x/.initial=.5,
  minor step y/.initial=.2,
  major step x/.initial=1,
  major step y/.initial=1,
  major line width/.initial=1pt,
}
\makeatother

\newlength\fheight
\newlength\fwidth

\usepackage[capitalize]{cleveref}
\crefname{section}{Sec.}{Secs.}

\IEEEoverridecommandlockouts

\NewSpotColorSpace{PANTONE}
\AddSpotColor{PANTONE} {PANTONE3015C} {PANTONE\SpotSpace 3015\SpotSpace C} {1 0.3 0 0.2}
\SetPageColorSpace{PANTONE}%

\usepackage{xcolor}

\usepackage{glossaries}
\makeglossaries
\glstoctrue

\newacronym{3dof}{3DoF}{3 Degrees of Freedom}
\newacronym{3gpp}{3GPP}{3rd Generation Partnership Project}
\newacronym{5g}{5G}{5\textsuperscript{th} Generation}
\newacronym{5gc}{5GC}{5G Core}
\newacronym{6dof}{6DoF}{6 Degrees of Freedom}
\newacronym{abft}{A-BFT}{Association-BeamForming Training}
\newacronym[firstplural=Access Categories (ACs)]{ac}{AC}{Access Category}
\newacronym{adc}{ADC}{Analog to Digital Converter}
\newacronym{addts}{ADDTS}{Add Traffic Stream}
\newacronym{afbw}{AFBW}{Average Fading Bandwidth}
\newacronym{aid}{AID}{Association ID}
\newacronym{aifs}{AIFS}{Arbitration Inter-Frame Space}
\newacronym{aimd}{AIMD}{Additive Increase Multiplicative Decrease}
\newacronym{am}{AM}{Acknowledged Mode}
\newacronym{amc}{AMC}{Adaptive Modulation and Coding}
\newacronym{ampdu}{A-MPDU}{MAC Protocol Data Unit Aggregation}
\newacronym{aoa}{AoA}{Angle of Arrival}
\newacronym{aod}{AoD}{Angle of Departure}
\newacronym{ap}{AP}{Access Point}
\newacronym{api}{API}{Application Programming Interface}
\newacronym{app}{APP}{Application}
\newacronym{aqm}{AQM}{Active Queue Management}
\newacronym{ar}{AR}{Augmented Reality}
\newacronym{arf}{ARF}{Auto Rate Fallback}
\newacronym{arp}{ARP}{Address Resolution Protocol}
\newacronym{ati}{ATI}{Announcement Transmission Interval}
\newacronym{awgn}{AGWN}{Additive White Gaussian Noise}
\newacronym{awv}{AWV}{Antenna Weight Vector}
\newacronym{balia}{BALIA}{Balanced Link Adaptation}
\newacronym{bdp}{BDP}{Bandwidth-Delay Product}
\newacronym{ber}{BER}{Bit Error Rate}
\newacronym{bf}{BF}{Beamforming}
\newacronym{bframe}{B-frame}{Bipredictive-coded frame}
\newacronym{bhi}{BHI}{Beacon Header Interval}
\newacronym{bi}{BI}{Beacon Interval}
\newacronym{brp}{BRP}{Beam Refinement Protocol}
\newacronym{bs}{BS}{Base Station}
\newacronym{bss}{BSS}{Basic Service Set}
\newacronym{bti}{BTI}{Beacon Transmission Interval}
\newacronym{cad}{CAD}{Computer-aided Design}
\newacronym{cbap}{CBAP}{Contention-Based Access Period}
\newacronym{cbr}{CBR}{Constant Bit Rate}
\newacronym{cc}{CC}{Congestion Control}
\newacronym{cdf}{CDF}{Cumulative Distribution Function}
\newacronym{cir}{CIR}{Channel Impulse Response}
\newacronym{cn}{CN}{Core Network}
\newacronym{cp}{CP}{Control Plane}
\newacronym{cqi}{CQI}{Channel Quality Indicator}
\newacronym{crs}{CRS}{Cell Reference Signal}
\newacronym{csirs}{CSI-RS}{Channel State Information - Reference Signal}
\newacronym{csmaca}{CSMA/CA}{Carrier Sense Multiple Access with Collision Avoidance}
\newacronym{cts}{CTS}{Clear to Send}
\newacronym{dc}{DC}{Dual Connectivity}
\newacronym{dce}{DCE}{Direct Code Execution}
\newacronym{dcf}{DCF}{Distributed Coordination Function}
\newacronym{dci}{DCI}{Downlink Control Information}
\newacronym{delts}{DELTS}{Delete Traffic Stream}
\newacronym{dl}{DL}{Downlink}
\newacronym{dmg}{DMG}{Directional Multi-Gigabit}
\newacronym{dmr}{DMR}{Deadline Miss Ratio}
\newacronym{dmrs}{DMRS}{DeModulation Reference Signal}
\newacronym{dti}{DTI}{Data Transmission Interval}
\newacronym{e2e}{E2E}{End-to-End}
\newacronym{ecn}{ECN}{Explicit Congestion Notification}
\newacronym{edca}{EDCA}{Enhanced Distributed Channel Access}
\newacronym{edf}{EDF}{Earliest Deadline First}
\newacronym{embb}{eMBB}{Enhanced Mobile BroadBand}
\newacronym{enb}{eNB}{evolved Node Base}
\newacronym{endc}{EN-DC}{E-UTRAN-\gls{nr} \gls{dc}}
\newacronym{epc}{EPC}{Evolved Packet Core}
\newacronym{es}{ES}{Edge Server}
\newacronym{ese}{ESE}{Extended Schedule Element}
\newacronym{fdd}{FDD}{Frequency Division Duplexing}
\newacronym{fdma}{FDMA}{Frequency Division Multiple Access}
\newacronym{fec}{FEC}{Forward Error Correction}
\newacronym{fov}{FoV}{Field-of-View}
\newacronym{fps}{FPS}{Frames Per Second}
\newacronym{fr2}{FR2}{Frequency Range 2}
\newacronym{fs}{FS}{Fast Switching}
\newacronym{ftp}{FTP}{File Transfer Protocol}
\newacronym{gmm}{GMM}{Gaussian Mixture Model}
\newacronym{gnb}{gNB}{Next Generation Node Base}
\newacronym[firstplural=Group of Pictures (GoPs)]{gop}{GoP}{Group of Pictures}
\newacronym{harq}{HARQ}{Hybrid Automatic Repeat reQuest}
\newacronym{hetnet}{HetNet}{Heterogeneous Network}
\newacronym{hh}{HH}{Hard Handover}
\newacronym{hmd}{HMD}{Head Mounted Device}
\newacronym{hol}{HOL}{Head-of-Line}
\newacronym{hqf}{HQF}{Highest-quality-first}
\newacronym{ia}{IA}{Initial Access}
\newacronym{iab}{IAB}{Integrated Access and Backhaul}
\newacronym{ibss}{IBSS}{Independent Basic Service Set}
\newacronym{id}{ID}{Identifier}
\newacronym{ifi}{IFI}{Inter-Frame Inter-arrival}
\newacronym{iframe}{I-frame}{Intra-coded frame}
\newacronym{imt}{IMT}{International Mobile Telecommunication}
\newacronym{imt2020}{IMT-2020}{International Mobile Te\-le\-com\-mu\-ni\-ca\-tion-2020}
\newacronym{inr}{INR}{Interference to Noise Ratio}
\newacronym{iot}{IoT}{Internet of Things}
\newacronym{ipa}{IPA}{Inter-Packet Arrival}
\newacronym{ism}{ISM}{Industrial, Scientific, and Medical}
\newacronym{itu}{ITU}{International Telecommunication Union}
\newacronym{kpi}{KPI}{Key Performance Indicator}
\newacronym{ks}{KS}{Kolmogorov-Smirnov}
\newacronym{lcf}{LCF}{Level Crossing Frequency}
\newacronym{lcm}{lcm}{least common multiple}
\newacronym{lcr}{LCR}{Level Crossing Rate}
\newacronym{los}{LoS}{Line-of-Sight}
\newacronym{lp}{LP}{Low Power}
\newacronym{lte}{LTE}{Long Term Evolution}
\newacronym{m2m}{M2M}{Machine to Machine}
\newacronym{mac}{MAC}{Medium Access Control}
\newacronym{mc}{MC}{Multi-Connectivity}
\newacronym{mcs}{MCS}{Modulation and Coding Scheme}
\newacronym{mec}{MEC}{Mobile Edge Cloud}
\newacronym{mi}{MI}{Mutual Information}
\newacronym{mib}{MIB}{Master Information Block}
\newacronym{mimo}{MIMO}{Multiple Input, Multiple Output}
\newacronym{ml}{ML}{Machine Learning}
\newacronym{mlr}{MLR}{Maximum-local-rate}
\newacronym[plural=\gls{mme}s,firstplural=Mobility Management Entities (MMEs)]{mme}{MME}{Mobility Management Entity}
\newacronym{mmw}{mmW}{Millimeter Wave}
\newacronym{moi}{MoI}{Method of Images}
\newacronym{mpc}{MPC}{Multi Path Component}
\newacronym{mpdu}{MPDU}{MAC Protocol Data Unit}
\newacronym{mptcp}{MPTCP}{Multipath TCP}
\newacronym{mr}{MR}{Mixed Reality}
\newacronym{mrdc}{MR-DC}{Multi \gls{rat} \gls{dc}}
\newacronym{msdu}{MSDU}{MAC Service Data Unit}
\newacronym{mss}{MSS}{Maximum Segment Size}
\newacronym{mtd}{MTD}{Machine-Type Device}
\newacronym{mtu}{MTU}{Maximum Transmission Unit}
\newacronym{mumimo}{MU-MIMO}{Multi-User Multiple Input, Multiple Output}
\newacronym{nav}{NAV}{Network Allocation Vector}
\newacronym{ncbr}{NCBR}{Non-Constant Bit Rate}
\newacronym{nfv}{NFV}{Network Function Virtualization}
\newacronym{nlos}{NLoS}{Non-Line-of-Sight}
\newacronym{nr}{NR}{New Radio}
\newacronym{nrmse}{NRMSE}{Normalized Root Mean Square Error}
\newacronym{ns2}{ns-2}{Network Simulator 2}
\newacronym{ns3}{ns-3}{Network Simulator 3}
\newacronym{nsa}{NSA}{Non Stand Alone}
\newacronym{o2i}{O2I}{Outdoor-to-Indoor}
\newacronym{ofdm}{OFDM}{Orthogonal Frequency Division Multiplexing}
\newacronym{pa}{PA}{Position-aware}
\newacronym{pan}{PAN}{Personal Area Network}
\newacronym{pbch}{PBCH}{Physical Broadcast Channel}
\newacronym{pbss}{PBSS}{Personal Basic Service Set}
\newacronym{pcf}{PCF}{Point Coordinator Function}
\newacronym{pcp}{PCP}{\gls{pbss} Central Point}
\newacronym{pcpap}{PCP/AP}{\acrlong{pcp}/\acrlong{ap}}
\newacronym{pdcch}{PDCCH}{Physical Downlonk Control Channel}
\newacronym{pdcp}{PDCP}{Packet Data Convergence Protocol}
\newacronym{pdf}{PDF}{Probability Density Function}
\newacronym{pdsch}{PDSCH}{Physical Downlink Shared Channel}
\newacronym{pdu}{PDU}{Packet Data Unit}
\newacronym{per}{PER}{Packet Error Rate}
\newacronym{pf}{PF}{Proportional Fair}
\newacronym{pframe}{P-frame}{Predictive-coded frame}
\newacronym{pgw}{PGW}{Packet Gateway}
\newacronym{phy}{PHY}{Physical Layer}
\newacronym{ppdu}{PPDU}{PHY Protocol Data Unit}
\newacronym{ppp}{PPP}{Poisson Point Process}
\newacronym{prb}{PRB}{Physical Resource Block}
\newacronym{pss}{PSS}{Primary Synchronization Signal}
\newacronym{pucch}{PUCCH}{Physical Uplink Control Channel}
\newacronym{pusch}{PUSCH}{Physical Uplink Shared Channel}
\newacronym{qd}{QD}{Quasi Deterministic}
\newacronym{qoe}{QoE}{Quality of Experience}
\newacronym{qos}{QoS}{Quality of Service}
\newacronym{rach}{RACH}{Random Access Channel}
\newacronym{ran}{RAN}{Radio Access Network}
\newacronym[firstplural=Radio Access Technologies (RATs)]{rat}{RAT}{Radio Access Technology}
\newacronym{red}{RED}{Random Early Detection}
\newacronym{rf}{RF}{Radio Frequency}
\newacronym{rl}{RL}{Reinforcement Learning}
\newacronym{rlc}{RLC}{Radio Link Control}
\newacronym{rlf}{RLF}{Radio Link Failure}
\newacronym{rr}{RR}{Round Robin}
\newacronym{rrc}{RRC}{Radio Resource Control}
\newacronym{rrm}{RRM}{Radio Resource Management}
\newacronym{rs}{RS}{Remote Server}
\newacronym{rsrp}{RSRP}{Reference Signal Received Power}
\newacronym{rsrq}{RSRQ}{Reference Signal Received Quality}
\newacronym{rss}{RSS}{Received Signal Strength}
\newacronym{rssi}{RSSI}{Received Signal Strength Indicator}
\newacronym{rt}{RT}{Ray Tracer}
\newacronym{rts}{RTS}{Request to Send}
\newacronym{rtt}{RTT}{Round Trip Time}
\newacronym{rw}{RW}{Receive Window}
\newacronym{rx}{RX}{Receiver}
\newacronym{sa}{SA}{standalone}
\newacronym{sack}{SACK}{Selective Acknowledgment}
\newacronym{sap}{SAP}{Service Access Point}
\newacronym{sc}{SC}{Single Carrier}
\newacronym{sch}{SCH}{Secondary Cell Handover}
\newacronym{scm}{SCM}{Spatial Channel Model}
\newacronym{scoot}{SCOOT}{Split Cycle Offset Optimization Technique}
\newacronym{sdma}{SDMA}{Spatial Division Multiple Access}
\newacronym{sdr}{SDR}{Software Defined Radio}
\newacronym{semm}{SEMM}{SPCA-EDCA Mixed Mode}
\newacronym{si}{SI}{Study Item}
\newacronym{sib}{SIB}{Secondary Information Block}
\newacronym{sinr}{SINR}{Signal-to-Interference-plus-Noise Ratio}
\newacronym{sir}{SIR}{Signal-to-Interference Ratio}
\newacronym{sls}{SLS}{Sector-Level Sweep}
\newacronym{sm}{SM}{Saturation Mode}
\newacronym{snr}{SNR}{Signal-to-Noise Ratio}
\newacronym{son}{SON}{Self-Organizing Network}
\newacronym{sp}{SP}{Service Period}
\newacronym{spr}{SPR}{Service Period Request}
\newacronym{srs}{SRS}{Sounding Reference Signal}
\newacronym{ss}{SS}{Synchronization Signal}
\newacronym{sss}{SSS}{Secondary Synchronization Signal}
\newacronym{ssw}{SSW}{Sector Sweep}
\newacronym{sta}{STA}{Station}
\newacronym{stb}{STB}{Set Top Box}
\newacronym{tb}{TB}{Transport Block}
\newacronym{tbtt}{TBTT}{Target Beacon Transmission Time}
\newacronym[firstplural=Traffic Categories (TCs)]{tc}{TC}{Traffic Category}
\newacronym{tcp}{TCP}{Transmission Control Protocol}
\newacronym{tdd}{TDD}{Time Division Duplexing}
\newacronym{tdma}{TDMA}{Time Division Multiple Access}
\newacronym{tfl}{TfL}{Transport for London}
\newacronym{tgad}{TGad}{Task Group ad}
\newacronym{tgay}{TGay}{Task Group ay}
\newacronym{tm}{TM}{Transparent Mode}
\newacronym{trp}{TRP}{Transmitter Receiver Pair}
\newacronym{ts}{TS}{Traffic Stream}
\newacronym{tsconst}{TSCONST}{Traffic Scheduling Constraint}
\newacronym{tsf}{TSF}{Timing Synchronization Function}
\newacronym{tspec}{TSPEC}{Traffic Specification}
\newacronym{tti}{TTI}{Transmission Time Interval}
\newacronym{ttt}{TTT}{Time-to-Trigger}
\newacronym{tx}{TX}{Transmitter}
\newacronym[firstplural=Transmission Opportunities (TXOPs)]{txop}{TXOP}{Transmission Opportunity}
\newacronym{udp}{UDP}{User Datagram Protocol}
\newacronym{ue}{UE}{User Equipment}
\newacronym{ul}{UL}{Uplink}
\newacronym{um}{UM}{Unacknowledged Mode}
\newacronym{uma}{UMa}{Urban Macro}
\newacronym{uml}{UML}{Unified Modeling Language}
\newacronym{up}{UP}{User Priority}
\newacronym{utc}{UTC}{Urban Traffic Control}
\newacronym{vbr}{VBR}{Variable Bit Rate}
\newacronym{vm}{VM}{Virtual Machine}
\newacronym{vr}{VR}{Virtual Reality}
\newacronym{wbf}{WBF}{Wired Bias Function}
\newacronym{wf}{WF}{Wired-first}
\newacronym{wifi}{Wi-Fi}{Wireless Fidelity}
\newacronym{wigig}{WiGig}{Wireless Gigabit}
\newacronym{wlan}{WLAN}{Wireless Local Area Network}
\newacronym{xr}{XR}{eXtended Reality}

\def\BibTeX{{\rm B\kern-.05em{\sc i\kern-.025em b}\kern-.08em
    T\kern-.1667em\lower.7ex\hbox{E}\kern-.125emX}}
\begin{document}
\history{Date of publication xxxx 00, 0000, date of current version xxxx 00, 0000.}
\doi{}

\title{An Open Framework for Analyzing and Modeling XR Network Traffic}
\author{
    \uppercase{Mattia Lecci}, \IEEEmembership{Graduate Student Member, IEEE},
    \uppercase{Matteo Drago}, \IEEEmembership{Graduate Student Member, IEEE},
    \uppercase{Andrea Zanella}, \IEEEmembership{Senior Member, IEEE}, and
    \uppercase{Michele Zorzi}, \IEEEmembership{Fellow, IEEE}}
\address{Department of Information Engineering, University of Padova, Italy (e-mails: \{leccimat, dragomat, zanella, zorzi\}@dei.unipd.it)}
\tfootnote{This work was partially supported by the National Institute of Standards and Technology (NIST) through Award No. 60NANB19D122.
    The identification of any commercial product or trade name does not imply endorsement or recommendation by NIST, nor is it intended to imply that the materials or equipment identified are necessarily the best available for the purpose.
    Mattia Lecci's activities were also supported by \textit{Fondazione CaRiPaRo} under the grant ``Dottorati di Ricerca 2018.''
    A preliminary version of this paper proposing a simpler traffic model with fewer traffic traces was presented in~\cite{lecci21bursty}.}

\markboth
{M. Lecci \headeretal: An Open Framework for Analyzing and Modeling XR Network Traffic}
{M. Lecci \headeretal: An Open Framework for Analyzing and Modeling XR Network Traffic}

\corresp{Corresponding author: Mattia Lecci (e-mail: leccimat@dei.unipd.it).}

\glsunset{wifi}

\begin{abstract}
    Thanks to recent advancements in the technology, \gls{xr} applications are gaining a lot of momentum, and they will surely become increasingly popular in the next decade.
    These new applications, however, require a step forward also in terms of models to simulate and analyze this type of traffic sources in modern communication networks, in order to guarantee to the users state of the art performance and \gls{qoe}.

    Recognizing this need, in this work, we present a novel open-source traffic model, which researchers can use as a starting point both for improvements of the model itself and for the design of optimized algorithms for the transmission of these peculiar data flows.
    Along with the mathematical model and the code, we also share with the community the traces that we gathered for our study, collected from freely available applications such as Minecraft VR, Google Earth VR, and Virus Popper.
    Finally, we propose a roadmap for the construction of an end-to-end framework that fills this gap in the current state of the art.
\end{abstract}


\titlepgskip=-15pt 

\maketitle

\begin{tikzpicture}[remember picture,overlay]
    \node[anchor=north,yshift=-5pt,xshift=-50pt] at (current page.north) {\fbox{\parbox{\dimexpr0.8\textwidth-\fboxsep-\fboxrule\relax}{
    \centering\footnotesize This paper has been submitted to IEEE Access. Copyright may change without notice.\\
    Please cite it as: M. Lecci, M. Drago, A. Zanella, M. Zorzi, "An Open Framework for Analyzing and Modeling XR Network Traffic," Submitted to IEEE Access, Pre-print available on arXiv.}}};
\end{tikzpicture}

\section{Introduction}
\label{sec:introduction}
\glsresetall
\glsunset{wifi}

After several years of innovations, the technology is finally ready for applications such as \gls{vr}, \gls{ar} and \gls{mr} to go mainstream (in the following we will use the term \gls{xr} as a general expression to consider all these distinct interaction modes).
According to some estimates, by 2025 there will be over 200 million people using \gls{xr} for immersive gaming experience and 95 million enjoying live events in this novel way~\cite{huaweiVrArWhitePaper}.
This immediately translates in an increasing sale of devices and headsets dedicated to experience this new type of contents, with an estimated shipment of these devices in the order of tens of millions in the coming decade, generating billions in revenue for all the fields in which this technology will be deployed~\cite{oculusVr,huaweiArInsight,pwcVr}.

Although it all started from the entertainment and video gaming arenas, where players could immerse in a virtual 3D world, now we can see \gls{xr} applied in various fields, such as building or landscape design, real estate, marketing, and healthcare, opening to the possibility of learning new concepts and training employees for difficult situations in a completely different way~\cite{zteVr,oculusVr,qualcommXr,ericsson2020,5gAmericasServicesInnovation,deloitte2018}.
Automotive companies, for example, are using \gls{vr} to cut the time that leads to the physical model of a new product from weeks to days~\cite{pwcVr}.
The peculiarity of this new class of contents, in fact, besides the wide range of use cases, is that the end user does not passively receive the information but, depending on the interaction with the virtual environment in which he/she is immersed, the traffic flow towards the content provider can change accordingly.
From a sales point of view, instead, \gls{vr} can give customers realistic experiences with products, allowing them to easily consider different options and configurations~\cite{oculusVr}.

In this paper, we will focus on examples related to the video gaming world (even though equivalent conclusions can be drawn for different \gls{xr} applications), where the user interacts with the application using a keyboard or joypad, and the results of such actions are immediately seen on the PC or TV screen.
Through \glspl{hmd}, when playing with videogames supporting \gls{vr}, users can also react by moving their heads and, for instance, depending on where the head is pointing, the application streams distinct portions of the environment~\cite{chiariotti2021vrSurvey}.

Even though traditionally gaming software ran on devices which needed to respect several hardware constraints to generate high-quality images, now the paradigm is shifting towards a cloud approach~\cite{nokiaCloudGaming,huawei2017cloudVr}. This can be extended to all other use cases besides gaming, and for this reason, we can refer to this new paradigm as \textit{Cloud XR}.
By moving the computing and graphical processing units into the cloud, in fact, less powerful devices can be used to fully exploit this new technology.
This would benefit not only in terms of the actual cost of an \gls{hmd} (which still plays a huge role in promoting the adoption by new users) but also in the final \gls{qoe}.
Having all the computing resources self-contained in the device would mean not only a higher weight and volume, but also concerns in terms of heat and battery life~\cite{huaweiArInsight, 5gAmericasServicesInnovation}.

This shift towards cloud infrastructures requires the optimization of current communication systems, to fully support distributed \gls{xr} services.
To this end, we need accurate models of the applications that generate these data flows and, to the best of our knowledge, no previous work has addressed this problem so far.

In this work we try to fill this gap by proposing a traffic model that emulates an \gls{xr} application, while also sketching a roadmap to guide researchers in the development of more precise models, using ours as a baseline.

To further understand what are the steps that most influence the \gls{xr} performance, it is useful to describe a common end-to-end \gls{xr} architecture~\cite{huaweiVrArWhitePaper, 5gAmericasServicesInnovation}.
First, we can start from the collection and processing of sensory and tracking information, delegated to an ad hoc device.
Then, this information is sent to an \gls{xr} server to compose the viewport, i.e., what is actually shown to the user.
This process includes 2D/3D media encoding and the generation of additional metadata (including the scene description).
The device's presentation engine at the client side, after receiving and decoding the information stream, generates the images to display.
These images are derived from the decoded signals, rendering metadata, and other information, if applicable.
Finally, video and audio tracks associated with the current pose are generated by synchronizing and spatially aligning the rendered media.
These steps need to be accomplished with minimal delay to guarantee adequate \gls{qoe}. 

In fact, the \textit{motion-to-photon} latency, i.e., the time from an action (e.g., a head movement) to the update of what is shown on the display must be below 20~ms to avoid the so called \textit{cybersickness}, associated to disorientation and dizziness~\cite{3gpp.26.928,huaweiVrArWhitePaper,hettinger1992motionSickness,groen2008motionSickness,vonMammen2016cyberSick,kim2017vrSickness}.
Following this physiological constraint, several industry players pose the network requirements, in terms of latency, in the range of 5--10~ms~\cite{3gpp.26.928,itu-t-f.743.10,huaweiVrArWhitePaper,huaweiArInsight,5gAmericasServicesInnovation}.
Also in terms of the gaming \gls{qoe}, it has been demonstrated that for first-person shooters, racing games, and team soccer matches, application latency directly impacts the results of competitive e-sports and, if not properly addressed, would lead to abandoning the game~\cite{nokiaCloudGaming}.
This translates into stringent constraints both in \gls{dl} and in \gls{ul}, considering that not only the content must be streamed as soon as it is required, but also the user movements need to be promptly notified to the server.
For this reason, the software that collects each movement input must consider all \gls{6dof}, tracking both translations and rotations in the three perpendicular axes (based on the \gls{vr} device, some may consider only rotational motion, i.e., \gls{3dof}). 
To take immersive mobile experience to the next level, many improvements will be required in head, body, and even gaze tracking~\cite{qualcommXr}.

It is also important to distinguish between processing latency, associated with computation and rendering, and network latency.
Rendering complex gaming images can be quite demanding, and the delay introduced by these operations can be larger than that caused by network services, which further motivates the need to offload these functions to proper cloud infrastructures~\cite{nokiaCloudGaming}.

Besides delay-related issues, an additional problem consists in the bursty nature of the \gls{xr} traffic, meaning that the throughput measured over short time windows could be much higher than its average value~\cite{5gAmericasServicesInnovation}, which can be the case for an application that periodically generates collections of packets to refresh the viewport.
Another aspect impacting the throughput is that, in order for the technology to be as close as possible to human vision, we will need a higher spatial and temporal resolution of the content presented to the user than currently possible (i.e., 3D 360\textdegree{} 8196$\times$4096 resolution at 90~Hz and beyond display refresh rate)~\cite{3gpp.26.928, qualcommXr, 5gAmericasServicesInnovation}.

The core technology that is expected to guarantee the satisfaction of all these requirements, by paving the way for an optimized distribution of processing capabilities, is 5G.
Many players have already invested in 5G for the rising of \gls{xr}, for operations at both sub-6~GHz and mmWave~\cite{ericsson2020,qualcommXr,orangeXR,samsung6gVision}.

Nonetheless, even though some efforts have also been devoted by standard bodies to the redaction of technical reports~\cite{3gpp.26.928, itu-t-f.743.10}, at the present time researchers are limited by the lack of precise traffic models representing the stream to/from an \gls{xr} server.
Having these models would allow the research community to design telecommunication solutions that could reduce the delay contribution related to the network, while also considering all the processing steps.
For this reason, we propose a generative model for \gls{xr} traffic sources, obtained from real application traces, and we also delineate a roadmap of the necessary steps to further improve it with additional features able to cope with the aforementioned problems, i.e., motion-to-photon latency, burstiness, capacity.

While in \cref{sec:soa} we summarize the current state of the art in the \gls{xr} arena, \cref{sec:vr_traffic_acquisition_and_analysis} is devoted to the description of the acquisition setup that we used to collect about 70~GB of data for a total of more than 4~hours of traced traffic time using different VR applications, both from the hardware and from the software point of view.
We will also describe each of these applications and illustrate how we analyzed the dataset.
The model obtained from this analysis will be presented in \cref{sec:traffic_model}, and its end-to-end validation, along with some example use cases, are discussed in \cref{sec:simulation_results}.
Finally, in \cref{sec:xr_traffic_modeling_roadmap} we propose a roadmap to extend our baseline model with additional increasingly complex features, and \cref{sec:conclusions} concludes the paper.

\section{State of the Art}
\label{sec:soa}

A seminal conceptual model that describes the human and technical elements creating the participatory environments of virtual reality systems was proposed in~\cite{latta1994conceptual}, dating back to 1994.
This demonstrates that the interest in the definition of common models for the study of this framework started even before the technology was ready, or even invented.

Despite the research interest in this field, to the best of our knowledge little work has been done on the creation of generative traffic models in \gls{xr} contexts, while the focus was put on different aspects of the technology.
In particular, a huge effort has been devoted to the creation and validation of practical systems that use immersive technology to interact with the world in different ways.
An example can be found in~\cite{hentschel2009vrSimulation}, which describes a system for the interactive analysis of large datasets with time-dependent data, realized on a multi-processor parallel machine in order to guarantee a smooth user experience.
Instead, in~\cite{saad2018VRsystem} the authors developed a proof-of-concept system, combining Oculus Rift HMD and the Phantom Premium 1.5 High Force haptic device with the goal of demonstrating the feasibility of combining HMD and haptics in one system.
Also, XR solutions have been tested for purposes of architectural design~\cite{ergun2019architect} and for providing virtual performance instructions and feedback on users that want to play a real piano~\cite{guo2021mentor}.

From a more technical perspective, a complete overview of the latest developments on immersive and $360^{\circ}$ video streaming can be found in~\cite{chiariotti2021vrSurvey}, where the author aims at providing a complete overview on four of the most important challenges in this field, namely: omnidirectional video coding and compression, subjective and objective \gls{qoe} and the factors that can affect it, saliency measurement and \gls{fov} prediction, and adaptive streaming of immersive $360^{\circ}$ videos.
As stressed in~\cite{chiariotti2021vrSurvey}, finding a proper way to measure the user's \gls{qoe} may be difficult.
This is especially important with respect to the design of telecommunication infrastructures able to optimize the experience of the user, and to guarantee constant and stable service quality.

For this reason, a lot of effort has been devoted to creating network solutions for the maximization of the quality of the delivered content.
In~\cite{yang2019QoeResourceAl}, for example, the authors proposed a scheme for uplink delivery of tile-based VR video over cellular networks. In particular, they formulate a resource allocation problem as a frequency and time-dependent non-deterministic polynomial NP-hard problem, and propose three distinct algorithms to solve it.
Instead, in~\cite{teng2021qoeMIMO} the authors consider a QoE-driven transmission of VR $360^{\circ}$ contents in a multi-user massive MIMO wireless network.
Specifically, in this scenario multiple users in the cell are requesting the same content, and the goal is to optimize the reception of such information through a stable scheme for the transmission of the viewport tiles.
In this work, they also try to allocate the power in order to guarantee a consistent delivery rate for each stream.

The impact of latency on the overall experience of the user has been mentioned in~\cref{sec:introduction}, along with the importance of tracking the movements of the user in applications with strict delay requirements.
The authors of~\cite{perfecto2020taming} used a real VR head-tracking dataset to maximize the quality of the delivered video chunk under low-latency constraints.
In that case, a deep recurrent neural network was designed for the prediction of the users' \gls{fov} (allowing to cluster those with overlapping FoV) while information on the future content and the users' locations was used as input of a proactive physical-layer multicast transmission scheme.

A key solution to the latency problem would be to rely on the capabilities of 5G and edge cloud, exploiting what has been referred to as \textit{Cloud XR} in~\cref{sec:introduction}.
Indeed, in~\cite{krogfoss20quantifying} the authors demonstrated that 5G and edge cloud are necessary to sustain the requirements of applications such as \gls{vr} gaming.

All these solutions, however, lack a model capable of generating data flows that can easily be associated with a real XR application.
The approach of~\cite{yang2019QoeResourceAl} consisted in using 240~frames of each 8K 360$^{\circ}$ uncompressed video sequence available from~\cite{lui2017sjtu}.
In that case the author applied the HVEC Kvazaar encoding procedure, setting the frame rate to 25~\gls{fps} and the \gls{gop} size to 8, and using a constant tiling scheme, ideal for the purpose of their work.
Despite the high level of details implemented in such a model, the use of a trace-based flow is limiting per se, considering also the limited portion of the video that they selected.
Having an offline encoding strategy is another limit, that in our framework has been overcome by integrating the rendering server in the processing pipeline.

Also in~\cite{teng2021qoeMIMO}, the simulation setup from the point of view of the VR architecture was defined in order to highlight the features of the algorithms proposed by the authors, and the nature of the traffic flow (e.g., average frame size, inter and intra-frames correlation, inter-frame interval, etc.) was not taken into account.

Regarding the problem of tracking the movements of the users, in~\cite{perfecto2020taming} the authors fed the recurrent neural network with the 3DoF traces from~\cite{lo2017hmddataset}, tracking the pose of 50 different users watching a catalog of 10 HD 360$^{\circ}$ videos from YouTube (60 seconds long, 4K resolution, 30 FPS, \gls{fov} of $100^{\circ} \times 100^{\circ}$).
Having a generative model that creates such a dataset based on statistical studies on a collection of different traces would have greatly aided the training of the neural network used in~\cite{perfecto2020taming}.
Also, finding a dataset that represents well the problem that we want to solve is usually not feasible, and this may further limit the research outcomes.

As a consequence, our goal is to provide the community with a tool for the automatic generation of such traces.
A preliminary version of this work was proposed in~\cite{lecci21bursty}, and here we extend it with the acquisition of longer and more heterogeneous traces, that now include realistic interaction with several VR applications.
This extension also allowed a more detailed and thorough validation of the model.
Besides making both the model and the traces public, we also propose a possible roadmap for making the framework as complete and detailed as possible, highlighting the most important contributions that would benefit researchers aiming at the design of ad hoc network protocol optimizations for this new type of traffic sources.

\section{VR Traffic: Acquisition and Analysis}
\label{sec:vr_traffic_acquisition_and_analysis}

In this section, we describe our basic traffic modeling work.
Specifically, in \cref{sub:acquisition_setup} we describe our acquisition setup and the VR applications that we acquired, then in \cref{sub:traffic_analysis} we analyze the raw traffic traces, and the different streams composing them, both in terms of content and in terms of statistics.

\subsection{Acquisition Setup}
\label{sub:acquisition_setup}

For the rendering server, we used a desktop PC equipped with an Intel Core~i7 processor, 32~GB of RAM, and an NVIDIA GeForce RTX~2080~Ti graphics card.
For the headset, instead, we used an iPhone~XS enclosed in a VR cardboard, which allows a realistic interaction with the applications.
The two nodes were connected via \gls{wifi} to improve the user's freedom of movement, at the cost of a slightly less stable channel and of possible interference from other surrounding devices.

VR applications were thus run on the rendering server and streamed to the headset using the application \textit{RiftCat~2.0} (on the server), and \textit{VRidge} 2.7.7 (on the phone).\footnote{\url{riftcat.com/vridge}}
This setup allows the user to play VR games on the SteamVR platform for up to a maximum of 10~minutes continuously, enough to obtain traffic traces to be analyzed (note that this limit is given by the free version of VRidge, and is absent in the premium version).
Many settings can be tuned in this application, such as the display resolution, the frame rate (either 30~or 60~FPS), the target data rate (i.e., the data rate the application will try to consistently stream to the client, which can be set from 1~to 50~Mbps), the video encoder (NVIDIA NVENC was used), and the video compression standard (H.264 was chosen), among other advanced settings.

As opposed to~\cite{lecci21bursty}, we acquired traces while realistically interacting with available VR applications using mouse, keyboard, and head movements.
Our setup only allowed us to interact with \gls{3dof}, i.e., the user was seated and only head rotations were sensed.
To simplify the analysis of the traffic stream, audio was not activated.

For this purpose, we selected three popular VR applications targeting different types of interactions.
Specifically:
\begin{itemize}
    \item \textit{Minecraft}: an extremely popular game, with the mod \textit{Vivecraft} enabling room-scale or seated VR experiences. The user can explore the virtual environment by walking or swimming, and interact with the virtual world by cutting trees, digging holes, crafting tools, etc.
    \item \textit{Virus Popper}: during this fast-paced educational game, many cartoony-looking viruses swarm a virtual room, and the user has to attack them with cleaning tools for survival.
    \item \textit{Google Earth VR - Tour}: the VR version of Google Earth, allowing a user to explore the world with satellite imagery, 3D terrain of the entire globe, and 3D buildings in hundreds of cities around the world. The SteamVR application also enables tours, teleporting the user all around the world every few seconds.
    \item \textit{Google Earth VR - Cities}: in this case, a more interactive experience is yielded, allowing the user to fully explore cities or landmarks for as long as they want.
\end{itemize}
Please note that \textit{Google Earth VR} was used in two different ways, thus allowing us to analyze two different versions of the same application.

To capture streamed packets, we ran Wireshark, a popular open-source packet analyzer, on the rendering server.
The traffic analysis was performed at 30~and 60~FPS for target data rates of \{10, 20, 30, 40, 50\}~Mbps and for all 4 applications with a resolution of 1920$\times$1080, for a total of over 70~GB of PCAP traces and 4~hours of analyzed \gls{vr} traffic.
Our dataset containing the processed VR traffic traces can be found within our software and can be easily reused, as later described in \cref{sec:simulation_results}.

\subsection{Traffic Analysis}
\label{sub:traffic_analysis}

\begin{figure*}[t!]
    \setlength\fheight{0.7\columnwidth}
    \setlength\fwidth{2.0\columnwidth}
    \centering
%
%

\definecolor{color0}{rgb}{1,0.498039215686275,0.0549019607843137}
\definecolor{color1}{rgb}{0.172549019607843,0.627450980392157,0.172549019607843}
\definecolor{color2}{rgb}{0.580392156862745,0.403921568627451,0.741176470588235}
\definecolor{color3}{rgb}{0.549019607843137,0.337254901960784,0.294117647058824}
\definecolor{color4}{rgb}{0.12156862745098,0.466666666666667,0.705882352941177}
\definecolor{color5}{rgb}{0.83921568627451,0.152941176470588,0.156862745098039}







\begin{tikzpicture}
\pgfplotsset{every tick label/.append style={font=\scriptsize}}

\begin{axis}[%
width=0,
height=0,
at={(0,0)},
scale only axis,
xmin=0,
xmax=0,
xtick={},
ymin=0,
ymax=0,
ytick={},
axis background/.style={fill=white},
legend style={legend cell align=center, align=center, draw=white!15!black, font=\scriptsize, at={(0, 0)}, anchor=center, /tikz/every even column/.append style={column sep=2em}},
legend columns=4,
]

\addplot [semithick, color1, mark=triangle*, mark size=3, mark options={solid,rotate=180}, only marks]
table {%
0 0
};
\addlegendentry{Video Frame (DL)}

\addplot [semithick, color4, mark=triangle*, mark size=3, mark options={solid,rotate=180}, only marks]
table {%
0 0
};
\addlegendentry{Frame Feedback (DL)}

\addplot [semithick, color0, mark=triangle*, mark size=3, mark options={solid}, only marks]
table {%
0 0
};
\addlegendentry{Frame Feedback (UL)}

\addplot [semithick, color5, mark=triangle*, mark size=3, mark options={solid}, only marks]
table {%
0 0
};
\addlegendentry{Head Tracking (UL)}

\end{axis}
\end{tikzpicture}%

    \input{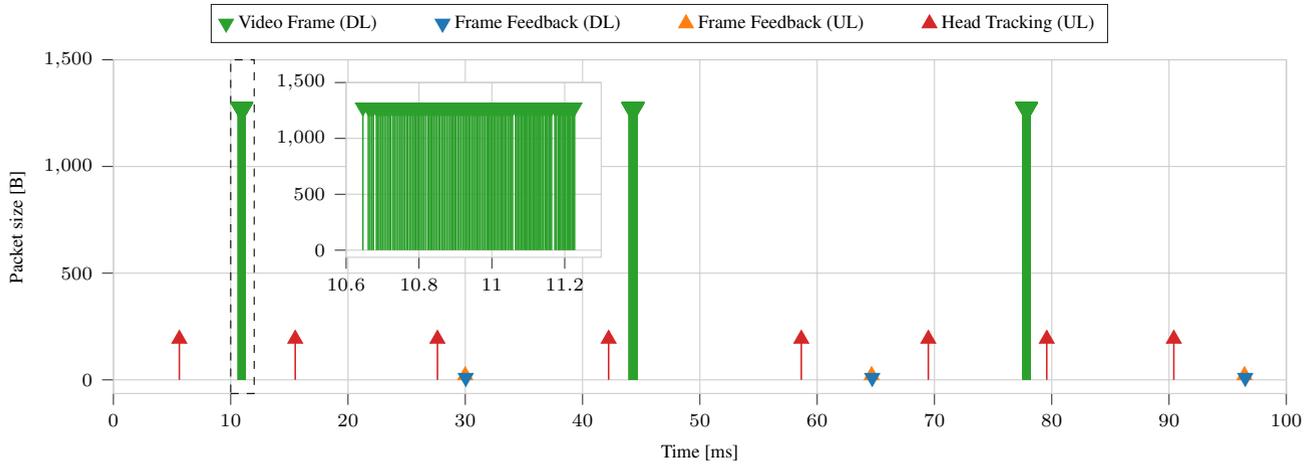}

    \caption{Portion of traffic trace from Virus Popper (50~Mbps, 30~FPS).
        For this trace, 130--140 individual fragments make up a video frame burst.}
    \label{fig:stem}
\end{figure*}

\begin{figure*}[t!]
    \setlength\fheight{0.55\columnwidth}
    \setlength\fwidth{0.7\columnwidth}
    \begin{subfigure}[t]{\textwidth}
        \centering
%
%

\definecolor{color0}{rgb}{0.12156862745098,0.466666666666667,0.705882352941177}
\definecolor{color1}{rgb}{1,0.498039215686275,0.0549019607843137}
\definecolor{color2}{rgb}{0.172549019607843,0.627450980392157,0.172549019607843}
\definecolor{color3}{rgb}{0.83921568627451,0.152941176470588,0.156862745098039}

\begin{tikzpicture}

\begin{axis}[%
width=0,
height=0,
at={(0,0)},
scale only axis,
xmin=0,
xmax=0,
xtick={},
ymin=0,
ymax=0,
ytick={},
axis background/.style={fill=white},
legend style={name=leg1, draw=none, 
            at={(0, 0)}, anchor=south, 
            inner ysep=0pt, 
            legend cell align=center, align=center, font=\scriptsize, /tikz/every even column/.append style={column sep=2em}},
legend columns=4,
]

\addlegendimage{only marks, color0}
\addlegendentry{Virus Popper}

\addlegendimage{only marks, color1}
\addlegendentry{Google Earth VR - Tour}

\addlegendimage{only marks, color2}
\addlegendentry{Google Earth VR - Cities}

\addlegendimage{only marks, color3}
\addlegendentry{Minecraft}

\end{axis}

\begin{axis}[%
    width=0,
    height=0,
    at={(0,0)},
    scale only axis,
    xmin=0,
    xmax=0,
    xtick={},
    ymin=0,
    ymax=0,
    ytick={},
    axis background/.style={fill=white},
    legend style={name=leg2, draw=none, 
                at={(0, 0)}, anchor=north, 
                inner ysep=0pt, 
                legend cell align=center, align=center, font=\scriptsize, /tikz/every even column/.append style={column sep=2em}},
    legend columns=4,
    ]
    
\addlegendimage {thick, black, dashed}
\addlegendentry{Expected}

\addlegendimage{only marks, black}
\addlegendentry{30 FPS}

\addlegendimage{only marks, black, mark=x, very thick, mark size=3}
\addlegendentry{60 FPS}

\end{axis}

\node [fit=(leg1)(leg2),draw, inner sep=0pt] {}; 

\end{tikzpicture}%
    \end{subfigure}
    \\
    \hspace*{\fill}%
    \begin{subfigure}[t]{0.3\textwidth}
        \centering
\begin{tikzpicture}

\definecolor{color0}{rgb}{0.12156862745098,0.466666666666667,0.705882352941177}
\definecolor{color1}{rgb}{1,0.498039215686275,0.0549019607843137}
\definecolor{color2}{rgb}{0.172549019607843,0.627450980392157,0.172549019607843}
\definecolor{color3}{rgb}{0.83921568627451,0.152941176470588,0.156862745098039}

\begin{axis}[
  width=\fwidth,
  height=\fheight,
legend cell align={left},
legend style={
  fill opacity=0.8,
  draw opacity=1,
  text opacity=1,
  at={(0.03,0.97)},
  anchor=north west,
  draw=white!80!black
},
tick align=outside,
tick pos=left,
x grid style={white!69.0196078431373!black},
xlabel={Target rate [Mbps]},
xmajorgrids,
xmin=8, xmax=52,
xtick style={color=black},
y grid style={white!69.0196078431373!black},
ylabel={Measured rate [Mbps]},
ymajorgrids,
ymin=-1, ymax=61.001394922243,
ytick style={color=black}
]
\addplot [thick, black, dashed, forget plot]
table {%
10 10
50 50
};
\addplot [only marks, color0, forget plot]
table {%
10 10.7781139037495
20 21.5735396873132
30 32.2345618757974
40 45.3638159923753
50 46.8967922425812
};
\addplot [only marks, color0, mark=x, very thick, mark size=3, forget plot]
table {%
10 11.2175737694826
20 21.5304679980273
30 32.3536374978889
40 43.1472517095796
50 53.7525348426884
};
\addplot [only marks, color1, forget plot]
table {%
10 9.71751340471715
20 21.3556004236347
30 32.1185376195354
40 42.8074269914892
50 58.5593053261703
};
\addplot [only marks, color1, mark=x, very thick, mark size=3, forget plot]
table {%
10 11.197751177045
20 21.3870962773907
30 32.0020221695676
40 42.429473309448
50 53.7191172718344
};
\addplot [only marks, color2, forget plot]
table {%
10 10.7902768223391
20 21.6029846837506
30 31.9670366745736
40 44.9341818853418
50 54.1884579430623
};
\addplot [only marks, color2, mark=x, very thick, mark size=3, forget plot]
table {%
10 11.2273249264578
20 21.5733343245202
30 32.3850990643272
40 43.1805473078194
50 53.8607937535441
};
\addplot [only marks, color3, forget plot]
table {%
10 10.7555288364806
20 21.0791800522904
30 31.5365821006962
40 42.2015993775022
50 51.2819500341559
};
\addplot [only marks, color3, mark=x, very thick, mark size=3, forget plot]
table {%
10 11.4568319780038
20 21.4796236686901
30 29.4208141529883
40 43.8668435877753
50 53.7623931205353
};
\end{axis}

\end{tikzpicture}
        \caption{Measured downlink data rate.}
        \label{fig:measured_rate_vs_data_rate}
    \end{subfigure}
    \hspace*{\fill}%
    \begin{subfigure}[t]{0.3\textwidth}
        \centering
\begin{tikzpicture}

\definecolor{color0}{rgb}{0.12156862745098,0.466666666666667,0.705882352941177}
\definecolor{color1}{rgb}{1,0.498039215686275,0.0549019607843137}
\definecolor{color2}{rgb}{0.172549019607843,0.627450980392157,0.172549019607843}
\definecolor{color3}{rgb}{0.83921568627451,0.152941176470588,0.156862745098039}

\begin{axis}[
  width=\fwidth,
  height=\fheight,
legend cell align={left},
legend style={
  fill opacity=0.8,
  draw opacity=1,
  text opacity=1,
  at={(0.03,0.97)},
  anchor=north west,
  draw=white!80!black
},
tick align=outside,
tick pos=left,
x grid style={white!69.0196078431373!black},
xlabel={Measured rate [Mbps]},
xmajorgrids,
xmin=7.27542380864449, xmax=61.001394922243,
xtick style={color=black},
y grid style={white!69.0196078431373!black},
ylabel={Avg. DL non-video rate [kbps]},
ymajorgrids,
ymin=-0.744789597213596, ymax=43.2818058236596,
ytick style={color=black}
]
\addplot [only marks,color0, forget plot]
table {%
10.7781139037495 2.70208031397651
21.5735396873132 2.70249216137055
32.2345618757974 2.71906984206224
45.3638159923753 3.32404256646038
46.8967922425812 2.96491642552847
};
\addplot [only marks,color0, mark=x, very thick, mark size=3, forget plot]
table {%
11.2175737694826 4.92634706497938
21.5304679980273 5.1175766766834
32.3536374978889 5.12424279426079
43.1472517095796 5.1440520956452
53.7525348426884 5.03269014186836
};
\addplot [only marks,color1, forget plot]
table {%
9.71751340471715 2.71327236084621
21.3556004236347 2.76670642589855
32.1185376195354 2.86272752267454
42.8074269914892 3.85995725674048
58.5593053261703 5.61765545414289
};
\addplot [only marks,color1, mark=x, very thick, mark size=3, forget plot]
table {%
11.197751177045 5.36198042677131
21.3870962773907 5.1102407859739
32.0020221695676 5.17357519161957
42.429473309448 5.16427233388763
53.7191172718344 5.21416774770934
};
\addplot [only marks,color2, forget plot]
table {%
10.7902768223391 2.70596324714662
21.6029846837506 2.70392346984171
31.9670366745736 2.76428378580091
44.9341818853418 3.06399481017632
54.1884579430623 4.65149859303789
};
\addplot [only marks,color2, mark=x, very thick, mark size=3, forget plot]
table {%
11.2273249264578 5.06876691708665
21.5733343245202 5.10616854385589
32.3850990643272 5.10346446302403
43.1805473078194 5.10885776301794
53.8607937535441 5.12419082479736
};
\addplot [only marks,color3, forget plot]
table {%
10.7555288364806 2.67829033477932
21.0791800522904 2.7216785949117
31.5365821006962 2.74308739083788
42.2015993775022 2.81029068681099
51.2819500341559 41.3483050860939
};
\addplot [only marks,color3, mark=x, very thick, mark size=3, forget plot]
table {%
11.4568319780038 5.54576537151659
21.4796236686901 5.33872072791733
29.4208141529883 5.09776082921428
43.8668435877753 5.30448091478905
53.7623931205353 5.1197152065749
};
\end{axis}

\end{tikzpicture}
        \caption{Average non-video DL rate.}
        \label{fig:avg_nonvideo_dl_rate}
    \end{subfigure}
    \hspace*{\fill}%
    \begin{subfigure}[t]{0.3\textwidth}
        \centering
\begin{tikzpicture}

\definecolor{color0}{rgb}{0.12156862745098,0.466666666666667,0.705882352941177}
\definecolor{color1}{rgb}{1,0.498039215686275,0.0549019607843137}
\definecolor{color2}{rgb}{0.172549019607843,0.627450980392157,0.172549019607843}
\definecolor{color3}{rgb}{0.83921568627451,0.152941176470588,0.156862745098039}

\begin{axis}[
  width=\fwidth,
  height=\fheight,
legend cell align={left},
legend style={
  fill opacity=0.8,
  draw opacity=1,
  text opacity=1,
  at={(0.03,0.03)},
  anchor=south west,
  draw=white!80!black
},
tick align=outside,
tick pos=left,
x grid style={white!69.0196078431373!black},
xlabel={Measured rate [Mbps]},
xmajorgrids,
xmin=7.27542380864449, xmax=61.001394922243,
xtick style={color=black},
y grid style={white!69.0196078431373!black},
ylabel={Avg. UL rate [kbps]},
ymajorgrids,
ymin=66.9192812258007, ymax=151.783427804791,
ytick style={color=black}
]
\addplot [only marks,color0, forget plot]
table {%
10.7781139037495 136.382834636454
21.5735396873132 135.694071084805
32.2345618757974 136.335074203772
45.3638159923753 134.226176138191
46.8967922425812 134.838029970375
};
\addplot [only marks,color0, mark=x, very thick, mark size=3, forget plot]
table {%
11.2175737694826 144.858990316748
21.5304679980273 146.488053296732
32.3536374978889 145.872575064513
43.1472517095796 147.925966596655
53.7525348426884 142.157053995945
};
\addplot [only marks,color1, forget plot]
table {%
9.71751340471715 137.018957710602
21.3556004236347 134.495921695722
32.1185376195354 134.055368701178
42.8074269914892 70.7767424339366
58.5593053261703 135.073447661243
};
\addplot [only marks,color1, mark=x, very thick, mark size=3, forget plot]
table {%
11.197751177045 143.94147766696
21.3870962773907 146.006085391748
32.0020221695676 145.284407910564
42.429473309448 145.97493813376
53.7191172718344 147.656509801236
};
\addplot [only marks,color2, forget plot]
table {%
10.7902768223391 139.97065016481
21.6029846837506 135.221192157135
31.9670366745736 134.365577888994
44.9341818853418 138.082548256901
54.1884579430623 135.634045831768
};
\addplot [only marks,color2, mark=x, very thick, mark size=3, forget plot]
table {%
11.2273249264578 139.489353751686
21.5733343245202 140.540949277024
32.3850990643272 144.665515376315
43.1805473078194 145.527264597675
53.8607937535441 146.529064156564
};
\addplot [only marks,color3, forget plot]
table {%
10.7555288364806 135.574990245592
21.0791800522904 134.9635169804
31.5365821006962 134.847902389498
42.2015993775022 134.671110756269
51.2819500341559 139.100976222816
};
\addplot [only marks,color3, mark=x, very thick, mark size=3, forget plot]
table {%
11.4568319780038 140.741382586248
21.4796236686901 142.018670392687
29.4208141529883 142.464162250622
43.8668435877753 142.012658929081
53.7623931205353 143.003870145628
};
\end{axis}

\end{tikzpicture}
        \caption{Average UL rate.}
        \label{fig:avg_ul_rate}
    \end{subfigure}
    \hspace*{\fill}%
    \\\vspace{2ex}\\
    \hspace*{\fill}%
    \begin{subfigure}[t]{0.3\textwidth}
        \centering
\begin{tikzpicture}

\definecolor{color0}{rgb}{0.12156862745098,0.466666666666667,0.705882352941177}
\definecolor{color1}{rgb}{1,0.498039215686275,0.0549019607843137}
\definecolor{color2}{rgb}{0.172549019607843,0.627450980392157,0.172549019607843}
\definecolor{color3}{rgb}{0.83921568627451,0.152941176470588,0.156862745098039}

\begin{axis}[
width=\fwidth,
height=\fheight,
legend cell align={left},
legend style={
  fill opacity=0.8,
  draw opacity=1,
  text opacity=1,
  at={(0.03,0.97)},
  anchor=north west,
  draw=white!80!black
},
tick align=outside,
tick pos=left,
x grid style={white!69.0196078431373!black},
xlabel={Measured rate [Mbps]},
xmajorgrids,
xmin=-3, xmax=63,
xtick style={color=black},
y grid style={white!69.0196078431373!black},
ylabel={Avg. video frame size [kB]},
ymajorgrids,
ymin=-5, ymax=262.5,
ytick style={color=black}
]
\addplot [thick, black, dashed, forget plot]
table {%
0 0
60 125
};
\addplot [thick, black, dashed, forget plot]
table {%
0 0
60 250
};
\addplot [only marks, color0, forget plot]
table {%
10.7781139037495 44.9055376884422
21.5735396873132 89.8885307646609
32.2345618757974 134.30422512925
45.3638159923753 189.040485981308
46.8967922425812 195.426711382814
};
\addplot [only marks, color0, mark=x, very thick, mark size=3, forget plot]
table {%
11.2175737694826 23.3787929729563
21.5304679980273 44.8618245989305
32.3536374978889 67.4233066185241
43.1472517095796 89.9299011327859
53.7525348426884 111.99264667413
};
\addplot [only marks, color1, forget plot]
table {%
9.71751340471715 40.4902136981557
21.3556004236347 88.9829428190446
32.1185376195354 133.833738228045
42.8074269914892 178.36812920194
58.5593053261703 244.013424150136
};
\addplot [only marks, color1, mark=x, very thick, mark size=3, forget plot]
table {%
11.197751177045 23.3636766467066
21.3870962773907 44.5588658061287
32.0020221695676 66.6835019349352
42.429473309448 88.3991131862711
53.7191172718344 111.907314193548
};
\addplot [only marks, color2, forget plot]
table {%
10.7902768223391 44.9582312849162
21.6029846837506 90.0065268292683
31.9670366745736 133.189373333333
44.9341818853418 187.227794842027
54.1884579430623 225.779685221675
};
\addplot [only marks, color2, mark=x, very thick, mark size=3, forget plot]
table {%
11.2273249264578 23.3930202972091
21.5733343245202 44.9429323809524
32.3850990643272 67.4656719344643
43.1805473078194 89.9619205900666
53.8607937535441 112.283820754717
};
\addplot [only marks, color3, forget plot]
table {%
10.7555288364806 44.827379330697
21.0791800522904 87.8570371053565
31.5365821006962 131.449602045354
42.2015993775022 175.850307495195
51.2819500341559 214.660193526635
};
\addplot [only marks, color3, mark=x, very thick, mark size=3, forget plot]
table {%
11.4568319780038 23.9565474911418
21.4796236686901 44.9023134631466
29.4208141529883 61.3822573477621
43.8668435877753 91.8647881351277
53.7623931205353 112.403719669938
};
\end{axis}

\end{tikzpicture}
        \caption{Average video frame size.}
        \label{fig:frame_size_vs_data_rate}
    \end{subfigure}
    \hspace*{\fill}%
    \begin{subfigure}[t]{0.3\textwidth}
        \centering
\begin{tikzpicture}

\definecolor{color0}{rgb}{0.12156862745098,0.466666666666667,0.705882352941177}
\definecolor{color1}{rgb}{1,0.498039215686275,0.0549019607843137}
\definecolor{color2}{rgb}{0.172549019607843,0.627450980392157,0.172549019607843}
\definecolor{color3}{rgb}{0.83921568627451,0.152941176470588,0.156862745098039}

\begin{axis}[
  width=\fwidth,
  height=\fheight,
legend cell align={left},
legend style={
  fill opacity=0.8,
  draw opacity=1,
  text opacity=1,
  at={(0.91,0.5)},
  anchor=east,
  draw=white!80!black
},
tick align=outside,
tick pos=left,
x grid style={white!69.0196078431373!black},
xlabel={Measured rate [Mbps]},
xmajorgrids,
xmin=7.20338907495301, xmax=62.5141243297641,
xtick style={color=black},
y grid style={white!69.0196078431373!black},
ylabel={Avg. IFI [ms]},
ymajorgrids,
ymin=14, ymax=36,
ytick style={color=black}
]
\addplot [thick, black, dashed, forget plot]
table {%
10 33.3333333333333
60 33.3333333333333
};
\addplot [thick, black, dashed, forget plot]
table {%
10 16.6666666666667
60 16.6666666666667
};
\addplot [only marks, color0, forget plot]
table {%
10.7781139037495 33.3340080037227
21.5735396873132 33.3335257171957
32.2345618757974 33.3335839404622
45.3638159923753 33.3392256292531
46.8967922425812 33.3392599672781
};
\addplot [only marks, color0, mark=x, very thick, mark size=3, forget plot]
table {%
11.2175737694826 16.6734768321949
21.5304679980273 16.6696450577854
32.3536374978889 16.672123573409
43.1472517095796 16.6742536354194
53.7525348426884 16.6683676643884
};
\addplot [only marks, color1, forget plot]
table {%
9.71751340471715 33.3357304295116
21.3556004236347 33.3347459993724
32.1185376195354 33.3350377529839
42.8074269914892 33.3366661186914
58.5593053261703 33.3359421303918
};
\addplot [only marks, color1, mark=x, very thick, mark size=3, forget plot]
table {%
11.197751177045 16.6917693786271
21.3870962773907 16.6681660945434
32.0020221695676 16.6702299906439
42.429473309448 16.6678331395144
53.7191172718344 16.6673399290246
};
\addplot [only marks, color2, forget plot]
table {%
10.7902768223391 33.3339551168638
21.6029846837506 33.333076339326
31.9670366745736 33.3324957866469
44.9341818853418 33.3362766627965
54.1884579430623 33.3331163478118
};
\addplot [only marks, color2, mark=x, very thick, mark size=3, forget plot]
table {%
11.2273249264578 16.6689792786806
21.5733343245202 16.6669870363511
32.3850990643272 16.6669061411431
43.1805473078194 16.6677934189099
53.8607937535441 16.6788820666417
};
\addplot [only marks, color3, forget plot]
table {%
10.7555288364806 33.3435582575847
21.0791800522904 33.3442121470129
31.5365821006962 33.349007893274
42.2015993775022 33.3369014094533
51.2819500341559 33.4887106889963
};
\addplot [only marks, color3, mark=x, very thick, mark size=3, forget plot]
table {%
11.4568319780038 16.7291738012231
21.4796236686901 16.723699625677
29.4208141529883 16.6913729388788
43.8668435877753 16.7523363546473
53.7623931205353 16.726236186228
};
\end{axis}

\end{tikzpicture}
        \caption{Video \acrfull{ifi} time.}
        \label{fig:ifi_vs_data_rate}
    \end{subfigure}
    \hspace*{\fill}%

    \caption{Results from acquired VR traffic traces.}
    \label{fig:vr_traces_vs_rate}
\end{figure*}
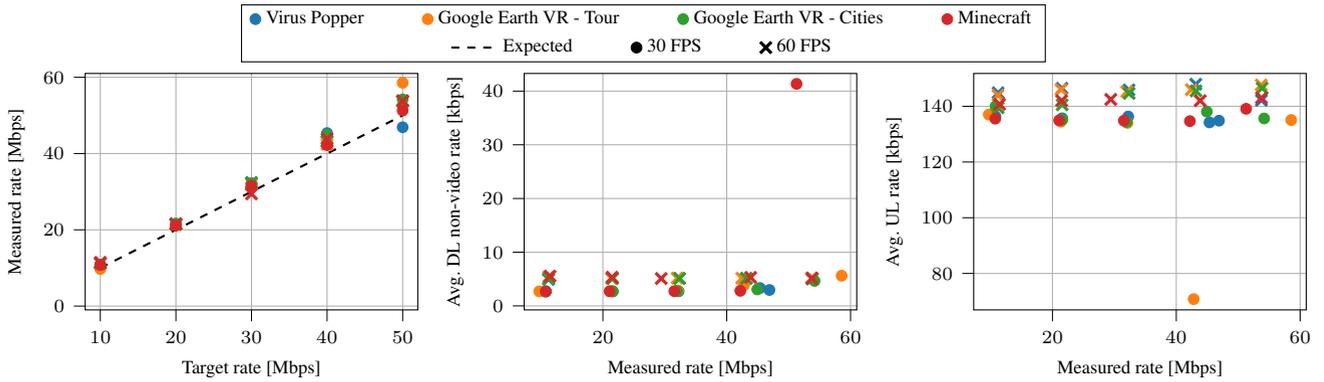
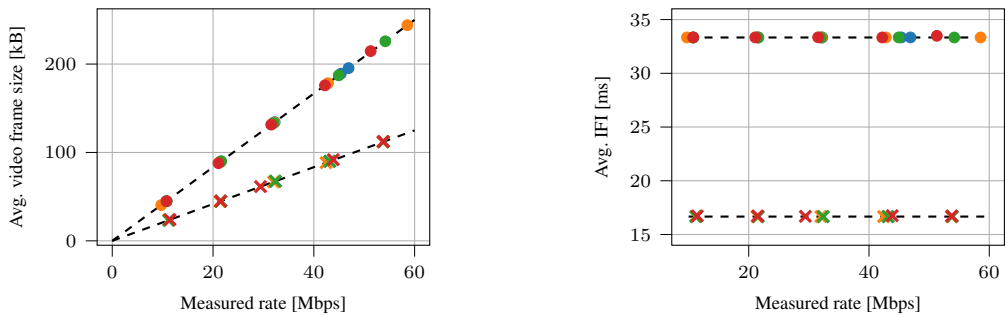


As described in~\cite{lecci21bursty}, we were able to partially reverse engineer both the \gls{dl} and \gls{ul} streams, and together with the help of \textit{RiftCat}'s developers, we are now able to reliably process the raw traffic traces.
We found that UDP sockets over IPv4 are used and both \gls{ul} and \gls{dl} streams contain several types of packets.
Specifically, the \gls{ul} stream contains packets such as synchronization, video frame reception information, and frequent small head-tracking information packets, whereas the \gls{dl} stream contains synchronization, acknowledgment, and video frame packet bursts.

To improve the stream quality, the RiftCat team developed a custom version of the ENet protocol\footnote{Available: \url{https://github.com/nxrighthere/ENet-CSharp}}, a relatively thin, simple and robust network communication layer on top of UDP, which offers reliable, in-order packet delivery.

In \cref{fig:stem} we show a visual representation of a slice of bidirectional VR streaming.
The plot shows the main data streams in both \gls{dl} and \gls{ul}, giving an idea of how this transmission works.

Most of the traffic is concentrated in \gls{dl} and is made up of packet bursts encoding video frames.
Video frame fragments were consistently found to be 1320~B long in all acquired traces, with a data size (the UDP payload) of 1278~B.
The last packet of the burst also has the same size as the others, suggesting that padding has been used in order to simplify the protocol, although this biases the frame size distribution to be discrete.

\begin{figure*}[t!]
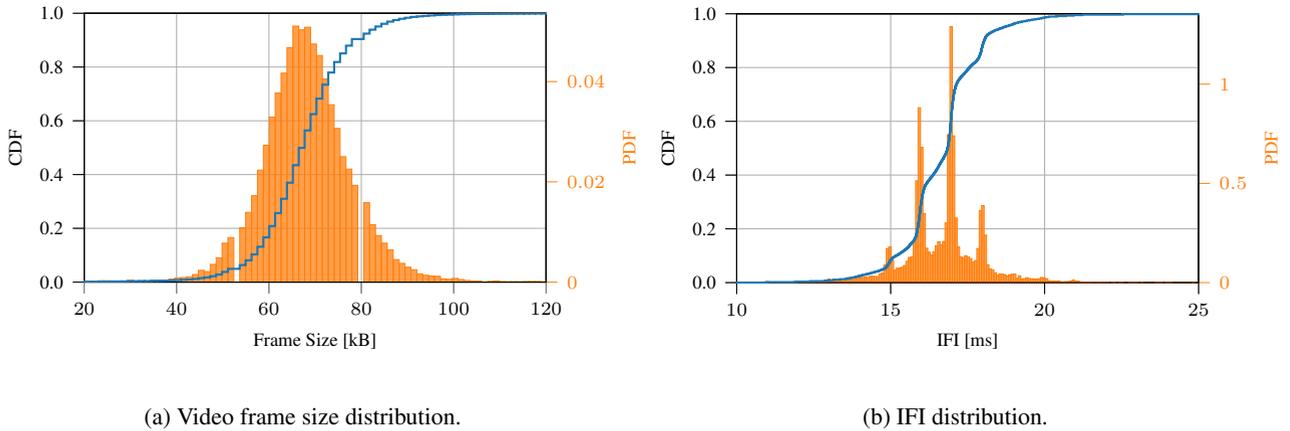

    \setlength\fheight{0.6\columnwidth}
    \setlength\fwidth{0.9\columnwidth}

    \hspace*{\fill}%
    \begin{subfigure}[t]{0.45\textwidth}
        \centering
        \input{img/frame_size_distrib.tikz}
        \caption{Video frame size distribution.}
        \label{fig:frame_size_distribution}
    \end{subfigure}
    \hspace*{\fill}%
    \begin{subfigure}[t]{0.45\textwidth}
        \centering
        \input{img/ifi_distrib.tikz}
        \caption{IFI distribution.}
        \label{fig:ifi_distribution}
    \end{subfigure}
    \hspace*{\fill}%

    \caption{Video frame distributions for Virus Popper (30~Mbps, 60~FPS).}
    \label{fig:distributions}
\end{figure*}

The second most noticeable traffic stream is the \gls{ul} head tracking information, which the headset acquires and sends to the rendering server to update the viewport to be rendered.
The head tracking payload was identified to be either 192~B or 97~B long, sometimes changing over the course of a single traffic trace, although the reason why different packet sizes were found is unclear.

Finally, smaller packets in both \gls{ul} and \gls{dl}, with payloads of respectively 21~B and 10~B, were identified to contain feedback on the reception of video frames, which is probably used in the streaming protocol to decide whether or not to retransmit some frames. 


By reverse-engineering the bits composing the UDP payload of video frames, it was possible to identify a recurring set of bits suggesting a 31~B APP-layer header and allowing us to identify some key fields, such as (i) the frame sequence number, (ii) the number of fragments composing the frame, (iii) the fragment sequence number, (iv) the total frame size, and (v) a checksum.
This information allowed us to reliably process and aggregate video frames.

Given the settings of the streaming application (i.e., frame rate and target data rate), it is clear that a \gls{cbr} video encoding is performed in the background.
In \cref{fig:measured_rate_vs_data_rate} we show the performance of the video encoder, almost always exceeding the target rate although by only 5--10\%.
A simple explanation of this behavior might be the underestimation of header sizes in the computations of the \gls{cbr} encoder, such as the header of the custom ENet protocol.
Notably, both frame rates behave similarly across all four applications, with stable performance.

\cref{fig:avg_nonvideo_dl_rate,fig:avg_ul_rate} show the low overhead due to non-video \gls{dl} and \gls{ul} transmissions (including head tracking), respectively.
Specifically, non-video \gls{dl} traffic only accounts for 3--5~kbps while \gls{ul} traffic for about 135--150~kbps, with 60~FPS traces consistently showing higher rates with respect to 30~FPS ones, probably due to the doubled amount of feedback.
Only two out of our forty traces show different rates, possibly due to some imperfection in the streaming.
In any case, these traffic flows are much lower than the target rates and appear constant, irrespective of the data rate or the application.
This consideration lead us to the decision of ignoring them, focusing only on modeling the \gls{dl} video traffic.

Considering $R$ the target data rate and $F$ the application frame rate, the \textit{average video frame size} is expected to be close to the ideal $S=R/F$, as shown in \cref{fig:frame_size_vs_data_rate}.
Note that the $x$-axis reports the measured data rate rather than the target data rate, i.e., the average data rate estimated from the acquired traces, which differs slightly from the target rate, as shown in \cref{fig:measured_rate_vs_data_rate}.

Furthermore, \cref{fig:ifi_vs_data_rate} shows that the average \gls{ifi} time perfectly matches the expected $1/F$, equal to $\mathrm{33.\bar{3}}$~ms for 30~FPS traces and $\mathrm{16.\bar{6}}$~ms for 60~FPS traces.

Moving to the analysis of the \glspl{pdf}, it is important to know that in a collection of packets associated to a video source, we can usually distinguish \glspl{iframe} (sometimes called called \textit{keyframes}), \glspl{pframe}, and \glspl{bframe}.
While \glspl{iframe} are compressed similarly to simple static pictures, \glspl{pframe} exploit the temporal correlation of successive frames to reduce the compressed frame size.
\glspl{bframe}, instead, can exploit the information from both previous and subsequent frames, further improving the compression efficiency at the cost of non-real-time transmission.
All the details associated with these compression techniques are regulated by standards like H.264~\cite{h264}.

Interestingly, a single clear peak is visible in \cref{fig:frame_size_distribution}, suggesting that different frames do not fall into different categories.
The \textit{RiftCat} team confirmed that no clear distinction among frame types is to be expected, since H.264~Periodic Intra Refresh is used when NVENC is selected as the encoder.
This means that instead of large keyframes, the image can be effectively refreshed over many frames using columns of intra-blocks that move across the video from one side to the other.
In this way, each frame can be capped to roughly the same size, which is extremely convenient to maintain a CBR stream.

As already mentioned, \textit{VRidge} simplifies the transmission by discretizing some units.
\cref{fig:frame_size_distribution} shows a clear staircase \gls{cdf} for the video frame size, suggesting that video frames have been padded as the underlying distribution is indeed discrete, with a distance between consecutive stairs of 1278~B, i.e., the UDP payload of packets containing fragments of video frames.

\begin{figure*}[t!]
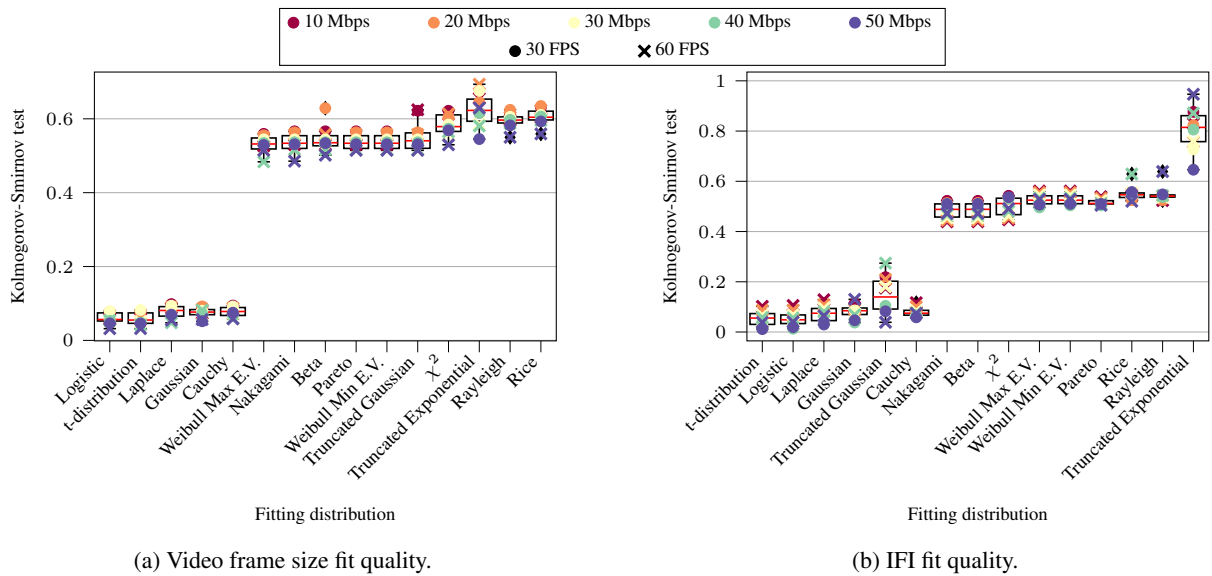

    \setlength\fheight{0.6\columnwidth}
    \setlength\fwidth{0.9\columnwidth}
    \begin{subfigure}[t]{\textwidth}
        \centering
%
%

\definecolor{color0}{rgb}{0.619607843137255,0.00392156862745098,0.258823529411765}
\definecolor{color1}{rgb}{0.974855824682814,0.557400999615532,0.322722029988466}
\definecolor{color2}{rgb}{0.998077662437524,0.99923106497501,0.746020761245675}
\definecolor{color3}{rgb}{0.527335640138408,0.810611303344867,0.645213379469435}
\definecolor{color4}{rgb}{0.368627450980392,0.309803921568627,0.635294117647059}

\begin{tikzpicture}

\begin{axis}[%
width=0,
height=0,
at={(0,0)},
scale only axis,
xmin=0,
xmax=0,
xtick={},
ymin=0,
ymax=0,
ytick={},
axis background/.style={fill=white},
legend style={name=leg1, draw=none, 
            at={(0, 0)}, anchor=south, 
            inner ysep=0pt, 
            legend cell align=center, align=center, font=\scriptsize, /tikz/every even column/.append style={column sep=2em}},
legend columns=5,
]

\addlegendimage{only marks, color0}
\addlegendentry{10 Mbps}

\addlegendimage{only marks, color1}
\addlegendentry{20 Mbps}

\addlegendimage{only marks, color2}
\addlegendentry{30 Mbps}

\addlegendimage{only marks, color3}
\addlegendentry{40 Mbps}

\addlegendimage{only marks, color4}
\addlegendentry{50 Mbps}

\end{axis}

\begin{axis}[%
    width=0,
    height=0,
    at={(0,0)},
    scale only axis,
    xmin=0,
    xmax=0,
    xtick={},
    ymin=0,
    ymax=0,
    ytick={},
    axis background/.style={fill=white},
    legend style={name=leg2, draw=none, 
                at={(0, 0)}, anchor=north, 
                inner ysep=0pt, 
                legend cell align=center, align=center, font=\scriptsize, /tikz/every even column/.append style={column sep=2em}},
    legend columns=5,
    ]
    
\addlegendimage{only marks, black}
\addlegendentry{30 FPS}

\addlegendimage{only marks, black, mark=x, very thick, mark size=3}
\addlegendentry{60 FPS}

\end{axis}

\node [fit=(leg1)(leg2),draw, inner sep=0pt] {}; 

\end{tikzpicture}%
    \end{subfigure}
    \\
    \hspace*{\fill}%
    \begin{subfigure}[t]{0.45\textwidth}
        \centering
        \input{img/boxplot_frame_size.tikz}
        \caption{Video frame size fit quality.}
        \label{fig:frame_size_boxplot}
    \end{subfigure}
    \hspace*{\fill}%
    \begin{subfigure}[t]{0.45\textwidth}
        \centering
        \input{img/boxplot_ifi.tikz}
        \caption{IFI fit quality.}
        \label{fig:ifi_boxplot}
    \end{subfigure}
    \hspace*{\fill}%

    \caption{Video frame fit qualities for the \textit{Google Earth VR - Cities} application.
        Fit quality is measured using the \acrshort{ks} test (lower is better).
        Box plots show median (red), 1\textsuperscript{st} and 3\textsuperscript{rd} quartiles (box), minimum and maximum (whiskers) of the \acrshort{ks} test with a given distribution, while markers show the exact values for the different traces.}
    \label{fig:boxplots}
\end{figure*}

Similarly, the \gls{ifi} time also appears to be discretized with a millisecond precision around the mean $\frac{1}{F}$, as seen in \cref{fig:ifi_distribution}, although some noise due to, e.g., variable rendering and encoding time, wireless channel condition, transmission queue state, transmission times, just to mention a few, smooths the CDF.

\section{Traffic Model}
\label{sec:traffic_model}


Following the analysis of \cref{sub:traffic_analysis}, in this section we will describe the proposed model for \gls{vr} traffic based on the collected VR traffic traces.

The analysis presented in the previous section reveals that both packet sizes and IFI times appear to be discrete in the collected data traces.
However, such granularity is likely due to specific design choices of the communication protocols used by the considered applications, rather than being a native characteristic of the XR services.
Therefore, we believe it is more suitable to use continuous random variables to model the size of the data blocks generated by the XR application and the time between them.
By doing so, we free our model from the specific constraints of this streaming application, with no loss of generality (as the discrete case can always be obtained from the continuous one), and in fact making it easier to accommodate other (non-discrete) cases in our framework if needed.

\subsection{Distribution fitting}
\label{sub:distribution_fitting}

Given the extremely large number of samples per trace (200--600~s at 30~or 60~FPS), common quality of fit statistical tests yield poor performance due to the discretized distributions.
Intuitively, while the \gls{pdf} of discrete and continuous distributions takes completely different forms, the \gls{cdf} of a discretized distribution is simply a staircased version of the related continuous distribution.
In that case, the goodness of fit can be tested by comparing the \glspl{cdf}, for example using the \gls{ks} test~\cite{kstest}, defined as:
\begin{equation}
    \mathrm{KS} = \sup_x \left| F_e(x) - F_t(x) \right|,
\end{equation}
where $\sup_x$ is the supremum of the set of distances, $F_e(x)$ is the empirical \gls{cdf} of the acquired data, and $F_t(x)$ is the \gls{cdf} of the target distribution.
The \gls{ks} test will thus be used to score the quality of fit, where values closer to zero indicate a better parameter estimation.

To fit and evaluate the best probability distributions for our data, we used the popular SciPy library~\cite{2020SciPy-NMeth}.
We tested 15 of the most common continuous univariate distributions available in the \texttt{scipy.stats} package, evaluating their performance on both frame size and IFI on our traffic traces.
Note that the SciPy library performs a maximum likelihood estimation of the parameters of the distribution, including \textit{location} and \textit{scale}, which SciPy adds to all continuous distributions by transforming the random variable $X$ into $(X - \mathrm{loc}) / \mathrm{scale}$.
Given the exceptional accordance between expected values and computed averages (see \cref{fig:frame_size_vs_data_rate,fig:ifi_vs_data_rate}) and considering the proposed generative model (described in \cref{sub:generative_model}), we fixed the location parameter to the expected value (i.e., $R/F$ for the frame size and $1/F$ for the IFI), fitting only the scale and the remaining parameters.
A selection of distributions is shown in \cref{fig:boxplots}.

\begin{figure*}[t!]
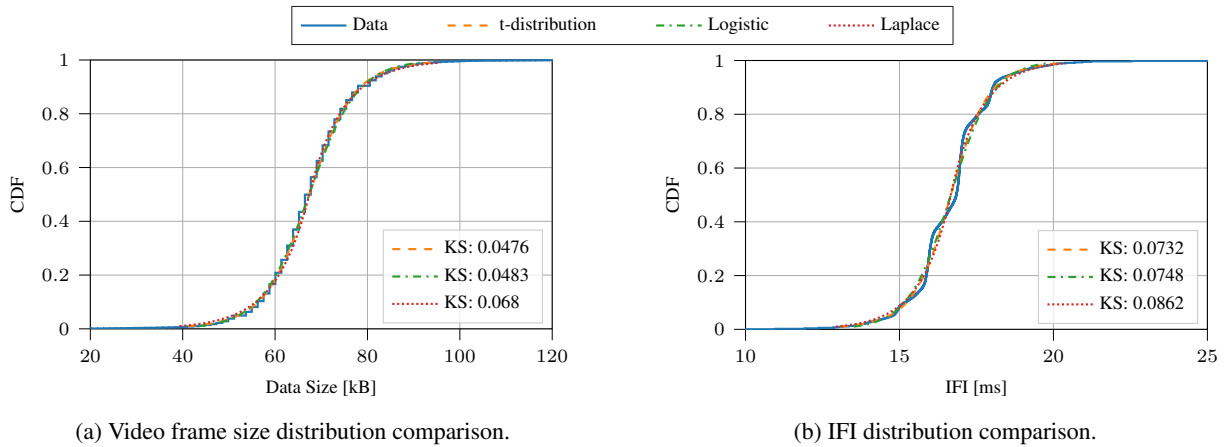

    \setlength\fheight{0.6\columnwidth}
    \setlength\fwidth{0.9\columnwidth}
    \begin{subfigure}[t]{\textwidth}
        \centering
%
%

\definecolor{color0}{rgb}{0.12156862745098,0.466666666666667,0.705882352941177}
\definecolor{color1}{rgb}{1,0.498039215686275,0.0549019607843137}
\definecolor{color2}{rgb}{0.172549019607843,0.627450980392157,0.172549019607843}
\definecolor{color3}{rgb}{0.83921568627451,0.152941176470588,0.156862745098039}
\definecolor{color4}{rgb}{0.580392156862745,0.403921568627451,0.741176470588235}

\begin{tikzpicture}
\pgfplotsset{every tick label/.append style={font=\scriptsize}}

\begin{axis}[%
width=0,
height=0,
at={(0,0)},
scale only axis,
xmin=0,
xmax=0,
xtick={},
ymin=0,
ymax=0,
ytick={},
axis background/.style={fill=white},
legend style={legend cell align=center, align=center, draw=white!15!black, font=\scriptsize, at={(0, 0)}, anchor=center, /tikz/every even column/.append style={column sep=2em}},
legend columns=5,
]

\addplot [thick, color0, const plot mark left]
table {%
0 0
};
\addlegendentry{Data}

\addplot [thick, color1, dashed]
table {%
0 0
};
\addlegendentry{t-distribution}

\addplot [thick, color2, dashdotted]
table {%
0 0
};
\addlegendentry{Logistic}

\addplot [thick, color3, densely dotted]
table {%
0 0
};
\addlegendentry{Laplace}

\end{axis}
\end{tikzpicture}%
    \end{subfigure}
    \\
    \hspace*{\fill}%
    \begin{subfigure}[t]{0.45\textwidth}
        \centering
        \input{img/fs_distrib_comparison.tikz}
        \caption{Video frame size distribution comparison.}
        \label{fig:fs_distrib_comparison}
    \end{subfigure}
    \hspace*{\fill}%
    \begin{subfigure}[t]{0.45\textwidth}
        \centering
        \input{img/ifi_distrib_comparison.tikz}
        \caption{IFI distribution comparison.}
        \label{fig:ifi_distrib_comparison}
    \end{subfigure}
    \hspace*{\fill}%

    \caption{Comparison of the three best fitting distributions for Virus Popper (30~Mbps, 60~FPS).
        The KS test is also shown, where lower values indicate a better fit.}
    \label{fig:distrib_comparison}
\end{figure*}

We found that the \textit{Student's~t} and \textit{Logistic} distributions, closely followed by the \textit{Laplace}, \textit{Gaussian}, and \textit{Cauchy} distributions, were the best fitting ones in almost all traces for both frame size and IFI.
\cref{fig:distrib_comparison} shows how similar the fitted distributions actually are.
Although the Student's~t distribution performs slightly better than the Logistic one in the slight majority of the collected traces, in our case the Logistic distribution was the best choice.
In fact, the third parameter of the Student's~t distribution is only able to yield minuscule improvements over the Logistic distribution, which only needs two parameters.
Furthermore, if custom simulators need to manually implement the desired random stream, the Student's~t distribution is very hard to reproduce~\cite{shaw2006student}, while the Logistic distribution requires a simple transformation.
This is the case when common libraries for random number generation cannot be used, such as in our implementation described in \cref{sec:simulation_results}.

As a reference, we use SciPy's definition of a logistic distribution, with \gls{pdf} in its standardized form as follows:
\begin{equation}
    f(x) = \frac{e^{-x}}{(1+e^{-x})^2}.
\end{equation}
To shift or scale the distribution, the location and scale parameters are used as previously described.



\subsection{Generative Model}
\label{sub:generative_model}

\begin{figure*}[t!]
    \setlength\fheight{0.6\columnwidth}
    \setlength\fwidth{0.9\columnwidth}
    \begin{subfigure}[t]{\textwidth}
        \centering
%
%

\definecolor{color0}{rgb}{0.12156862745098,0.466666666666667,0.705882352941177}
\definecolor{color1}{rgb}{1,0.498039215686275,0.0549019607843137}
\definecolor{color2}{rgb}{0.172549019607843,0.627450980392157,0.172549019607843}
\definecolor{color3}{rgb}{0.83921568627451,0.152941176470588,0.156862745098039}
\definecolor{color4}{rgb}{0.580392156862745,0.403921568627451,0.741176470588235}

\begin{tikzpicture}
\pgfplotsset{every tick label/.append style={font=\scriptsize}}

\begin{axis}[%
width=0,
height=0,
at={(0,0)},
scale only axis,
xmin=0,
xmax=0,
xtick={},
ymin=0,
ymax=0,
ytick={},
axis background/.style={fill=white},
legend style={legend cell align=center, align=center, draw=white!15!black, font=\scriptsize, at={(0, 0)}, anchor=center, /tikz/every even column/.append style={column sep=2em}},
legend columns=5,
]

\addplot [only marks, black]
table {%
0 0
};
\addlegendentry{30 FPS}

\addplot [only marks, black, mark=x, very thick, mark size=3]
table {%
0 0
};
\addlegendentry{60 FPS}

\addplot [thick, red, dashed]
table {%
0 0
};
\addlegendentry{Fit}

\end{axis}
\end{tikzpicture}%
    \end{subfigure}
    \\\\
    \hspace*{\fill}%
    \begin{subfigure}[t]{0.45\textwidth}
        \centering
\begin{tikzpicture}

\definecolor{color0}{rgb}{0.12156862745098,0.466666666666667,0.705882352941177}
\definecolor{color1}{rgb}{1,0.498039215686275,0.0549019607843137}

\begin{axis}[
    width=\fwidth,
    height=\fheight,
legend cell align={left},
legend style={fill opacity=0.8, draw opacity=1, text opacity=1, draw=white!80!black},
tick align=outside,
tick pos=left,
x grid style={white!69.0196078431373!black},
xlabel={Measured rate [Mbps]},
xmajorgrids,
xmin=-1.95, xmax=62.95,
xtick style={color=black},
y grid style={white!69.0196078431373!black},
ylabel={Fitted video frame size dispersion},
ymajorgrids,
ymin=0.075, ymax=0.165,
ytick style={color=black},
yticklabel style={
        /pgf/number format/fixed,
        /pgf/number format/precision=2
},
scaled y ticks=false
]
\addplot [only marks, black, forget plot]
table {%
10.7902768223391 0.123938705528296
21.6029846837506 0.123468300546448
31.9670366745736 0.0839159887710853
44.9341818853418 0.0971693162686376
54.1884579430623 0.0905979744440008
};
\addplot [only marks, black, mark=x, very thick, mark size=3, forget plot]
table {%
11.2273249264578 0.140981826533797
21.5733343245202 0.109135788172187
32.3850990643272 0.115025025329795
43.1805473078194 0.103963403896968
53.8607937535441 0.0938197319061051
};
\addplot [thick, red, dashed, forget plot]
table {%
1 0.259684566630138
1.5959595959596 0.23062088808887
2.19191919191919 0.212769995571987
2.78787878787879 0.200166120034649
3.38383838383838 0.190558537286394
3.97979797979798 0.182869262187692
4.57575757575758 0.176503693548238
5.17171717171717 0.171101345683517
5.76767676767677 0.166428272070855
6.36363636363636 0.162324620365508
6.95959595959596 0.158676661590483
7.55555555555556 0.155400802473477
8.15151515151515 0.152433928805197
8.74747474747475 0.14972730392401
9.34343434343434 0.147242566017472
9.93939393939394 0.144949017490791
10.5353535353535 0.142821738580713
11.1313131313131 0.140840243145045
11.7272727272727 0.138987500730409
12.3232323232323 0.137249211956915
12.9191919191919 0.135613262772785
13.5151515151515 0.134069307373538
14.1111111111111 0.132608445225
14.7070707070707 0.131222967954881
15.3030303030303 0.129906158831471
15.8989898989899 0.128652132317762
16.4949494949495 0.127455704515803
17.0909090909091 0.126312287671843
17.6868686868687 0.125217803604451
18.2828282828283 0.124168612148417
18.8787878787879 0.123161451613116
19.4747474747475 0.122193388928359
20.0707070707071 0.12126177765781
20.6666666666667 0.120364222445085
21.2626262626263 0.119498548752589
21.8585858585859 0.118662776981028
22.4545454545455 0.117855100234938
23.050505050505 0.117073865138762
23.6464646464646 0.116317555217947
24.2424242424242 0.11558477644693
24.8383838383838 0.114874244635842
25.4343434343434 0.114184774384031
26.030303030303 0.113515269374088
26.6262626262626 0.11286471381709
27.2222222222222 0.112232164890121
27.8181818181818 0.111616746032024
28.4141414141414 0.111017640983884
29.010101010101 0.11043408847783
29.6060606060606 0.109865377491877
30.2020202020202 0.109310843000446
30.7979797979798 0.108769862160118
31.3939393939394 0.108241850878576
31.989898989899 0.107726260721779
32.5858585858586 0.107222576120392
33.1818181818182 0.106730311841648
33.7777777777778 0.106249010697133
34.3737373737374 0.105778241460769
34.969696969697 0.10531759697442
35.5656565656566 0.104866692421348
36.1616161616162 0.10442516375011
36.7575757575758 0.103992666233536
37.3535353535353 0.103568873149244
37.9494949494949 0.103153474569666
38.5454545454545 0.102746176250948
39.1414141414141 0.102346698611244
39.7373737373737 0.101954775789984
40.3333333333333 0.101570154780571
40.9292929292929 0.101192594629808
41.5252525252525 0.100821865698008
42.1212121212121 0.100457748974414
42.7171717171717 0.100100035443053
43.3131313131313 0.0997485254946771
43.9090909090909 0.0994030283808558
44.5050505050505 0.0990633617066702
45.1010101010101 0.098729350958803
45.6969696969697 0.0984008290661234
46.2929292929293 0.0980776359901393
46.8888888888889 0.0977596183429293
47.4848484848485 0.0974466290303883
48.0808080808081 0.0971385269188149
48.6767676767677 0.0968351765230449
49.2727272727273 0.0965364477144951
49.8686868686869 0.0962422154476191
50.4646464646465 0.0959523595034138
51.0606060606061 0.0956667642487225
51.6565656565657 0.0953853184101921
52.2525252525252 0.0951079148618339
52.8484848484848 0.0948344504252243
53.4444444444444 0.0945648256814614
54.040404040404 0.0942989447940611
54.6363636363636 0.0940367153420458
55.2323232323232 0.0937780481625335
55.8282828282828 0.0935228572021907
56.4242424242424 0.093271059376962
57.020202020202 0.0930225744395324
57.6161616161616 0.0927773248540209
58.2121212121212 0.0925352356774398
58.8080808080808 0.0922962344474895
59.4040404040404 0.0920602510762908
60 0.0918272177496828
};
\end{axis}

\end{tikzpicture}
        \caption{Video frame size model.}
        \label{fig:frame_size_model}
    \end{subfigure}
    \hspace*{\fill}%
    \begin{subfigure}[t]{0.45\textwidth}
        \centering
\begin{tikzpicture}

\definecolor{color0}{rgb}{0.12156862745098,0.466666666666667,0.705882352941177}
\definecolor{color1}{rgb}{1,0.498039215686275,0.0549019607843137}
\definecolor{color2}{rgb}{0.172549019607843,0.627450980392157,0.172549019607843}

\begin{axis}[
  width=\fwidth,
  height=\fheight,
legend cell align={left},
legend style={
  fill opacity=0.8,
  draw opacity=1,
  text opacity=1,
  at={(0.97,0.03)},
  anchor=south east,
  draw=white!80!black
},
tick align=outside,
tick pos=left,
x grid style={white!69.0196078431373!black},
xlabel={Measured rate [Mbps]},
xmajorgrids,
xmin=-1.95, xmax=62.95,
xtick style={color=black},
y grid style={white!69.0196078431373!black},
ylabel={Fitted IFI dispersion},
ymajorgrids,
ymin=0.018, ymax=0.041,
ytick style={color=black},
yticklabel style={
        /pgf/number format/fixed,
        /pgf/number format/precision=2
},
scaled y ticks=false
]
\addplot [only marks, black, forget plot]
table {%
10.7902768223391 0.0200130188661049
21.6029846837506 0.0219833260482765
31.9670366745736 0.0247175532656961
44.9341818853418 0.0276464187417084
54.1884579430623 0.0351446896490849
};
\addplot [only marks, black, mark=x, very thick, mark size=3, forget plot]
table {%
11.2273249264578 0.031664492377371
21.5733343245202 0.0345229583217548
32.3850990643272 0.0380045234721795
43.1805473078194 0.0352408522199022
53.8607937535441 0.0334254546218456
};
\addplot [thick, red, dashed, forget plot]
table {%
1 0.0345716562026106
60 0.0345716562026106
};
\addplot [thick, red, dashed, forget plot]
table {%
1 0.00895303711694265
1.5959595959596 0.0103584453305591
2.19191919191919 0.0114360581250393
2.78787878787879 0.0123269309253429
3.38383838383838 0.0130947612647012
3.97979797979798 0.0137743768043914
4.57575757575758 0.0143871315219419
5.17171717171717 0.0149471681946896
5.76767676767677 0.015464390625749
6.36363636363636 0.0159460336695819
6.95959595959596 0.0163975609388446
7.55555555555556 0.0168232105082577
8.15151515151515 0.0172263432552913
8.74747474747475 0.0176096742849106
9.34343434343434 0.0179754318905571
9.93939393939394 0.0183254698684689
10.5353535353535 0.018661348821025
11.1313131313131 0.0189843962634023
11.7272727272727 0.0192957518872172
12.3232323232323 0.0195964022070227
12.9191919191919 0.01988720746799
13.5151515151515 0.0201689228171408
14.1111111111111 0.0204422151577005
14.7070707070707 0.0207076767102319
15.3030303030303 0.0209658360301722
15.8989898989899 0.0212171670384503
16.4949494949495 0.0214620964838995
17.0909090909091 0.0217010101561221
17.6868686868687 0.0219342580939452
18.2828282828283 0.0221621589799375
18.8787878787879 0.022385003870355
19.4747474747475 0.0226030593786555
20.0707070707071 0.0228165704067761
20.6666666666667 0.0230257624998334
21.2626262626263 0.023230843885449
21.8585858585859 0.0234320072475304
22.4545454545455 0.0236294312753291
23.050505050505 0.0238232820214127
23.6464646464646 0.0240137140964198
24.2424242424242 0.0242008717238065
24.8383838383838 0.0243848896740059
25.4343434343434 0.0245658940943317
26.030303030303 0.0247440032484153
26.6262626262626 0.0249193281768699
27.2222222222222 0.0250919732891394
27.8181818181818 0.025262036895037
28.4141414141414 0.0254296116832761
29.010101010101 0.0255947851532698
29.6060606060606 0.025757640005627
30.2020202020202 0.0259182544960398
30.7979797979798 0.0260767027566478
31.3939393939394 0.0262330550884332
31.989898989899 0.0263873782277587
32.5858585858586 0.0265397355897684
33.1818181818182 0.0266901874910463
33.7777777777778 0.0268387913536381
34.3737373737374 0.0269856018922954
34.969696969697 0.0271306712865888
35.5656565656566 0.0272740493393467
36.1616161616162 0.0274157836227174
36.7575757575758 0.027555919613008
37.3535353535353 0.0276945008153282
37.9494949494949 0.0278315688789628
38.5454545454545 0.027967163704292
39.1414141414141 0.0281013235420037
39.7373737373737 0.0282340850852586
40.3333333333333 0.028365483555408
40.9292929292929 0.0284955527818033
41.5252525252525 0.0286243252761835
42.1212121212121 0.028751832302082
42.7171717171717 0.0288781039396511
43.3131313131313 0.0290031691462668
43.9090909090909 0.0291270558132418
44.5050505050505 0.0292497908189466
45.1010101010101 0.0293714000786103
45.6969696969697 0.0294919085910501
46.2929292929293 0.0296113404825567
46.8888888888889 0.0297297190481418
47.4848484848485 0.0298470667903386
48.0808080808081 0.0299634054557287
48.6767676767677 0.0300787560693561
49.2727272727273 0.0301931389671729
49.8686868686869 0.0303065738266531
50.4646464646465 0.0304190796956988
51.0606060606061 0.0305306750199513
51.6565656565657 0.0306413776686141
52.2525252525252 0.0307512049588853
52.8484848484848 0.0308601736790872
53.4444444444444 0.0309683001105791
54.040404040404 0.0310756000485281
54.6363636363636 0.0311820888216102
55.2323232323232 0.0312877813107075
55.8282828282828 0.0313926919666636
56.4242424242424 0.031496834827153
57.020202020202 0.0316002235327181
57.6161616161616 0.0317028713420239
58.2121212121212 0.0318047911463749
58.8080808080808 0.0319059954835376
59.4040404040404 0.0320064965509092
60 0.0321063062180683
};

\node (60fps) at (10,0.038) {\scriptsize 60 FPS};
\draw[->] (60fps) edge (10, 0.035);

\node[anchor=north] (30fps) at (55.2323232323232, 0.026) {\scriptsize 30 FPS};
\draw[->] (30fps) edge (55.2323232323232, 0.031);

\end{axis}

\end{tikzpicture}
        \caption{IFI model.}
        \label{fig:ifi_model}
    \end{subfigure}
    \hspace*{\fill}%

    \caption{Generalization models for the \textit{Google Earth VR - Cities} application.
        Individual points show the \textit{scale} parameter of the Logistic model fitted on the acquired data, while the dashed red lines attempt to generalize the model to intermediate target data rates.}
    \label{fig:models}
\end{figure*}
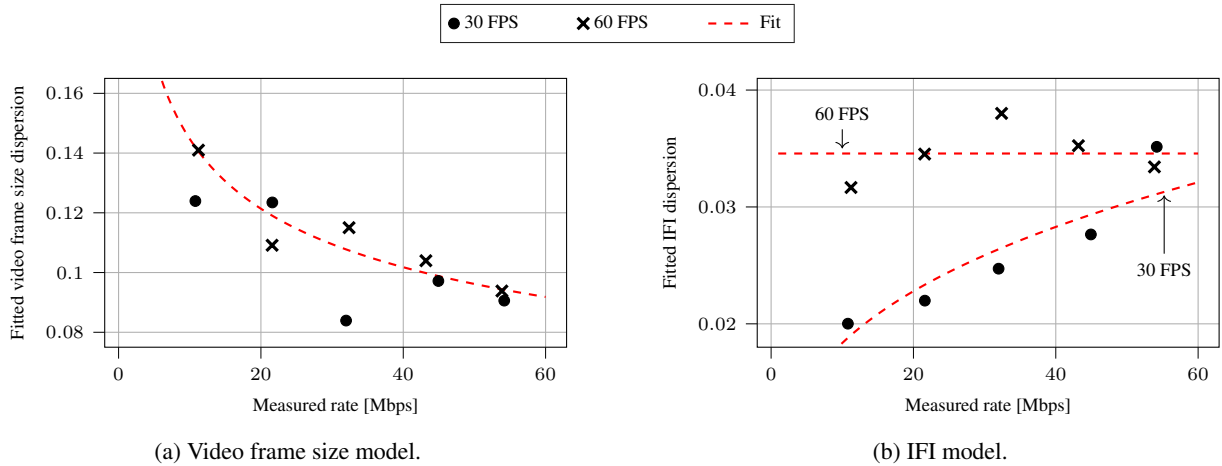

Now that we characterized and fitted the statistical distributions of the 40 acquired traces, we want to define a generative model which would allow a user to synthesize XR traffic at will, be it for analysis or simulation purposes.
As already discussed in \cref{sub:first_order_stats}, in this paper we propose a simple generative model, that only attempts to capture the statistical distributions of video frame size and \acrfullpl{ifi}, leaving higher-order statistical descriptions for future work.

We define the \textit{dispersion} as the ratio of the \textit{scale} over the \textit{location} parameter, attempting to find a common value for both frame rates, since absolute values are likely to differ by a constant factor (see \cref{fig:frame_size_vs_data_rate,fig:ifi_vs_data_rate}).
While data aggregation is doable for frame sizes (as shown, for example, in \cref{fig:frame_size_model}), data for IFI did not allow us to do so.
As shown in \cref{fig:ifi_model}, in fact, data for 30~and 60~FPS behaves differently, making it impossible for us to create a single model for this parameter.
This implies that our model will only be able to generalize over data rates, whereas 30~and 60~FPS are the only supported frame rates, and modeling and testing different values would require new data for the corresponding frame rate.

After carefully studying the acquired traffic traces, we propose to generalize the scale parameters for both video frame size and \gls{ifi} time with a power law, namely:
\begin{equation}\label{eq:power_law}
    y = a x^b.
\end{equation}
Furthermore, as \cref{fig:ifi_model} suggests, the 60~FPS IFI fits for all applications resulted in $|b|<10^{-4}$, suggesting a constant behavior, irrespective of the data rate.
In that case, we thus assumed a constant fit (a corner case of power law with $b=0$) by computing the average value across all tested target data rates.

As can clearly be observed from the collected data, the proposed model has been extracted from acquisitions between about 10~and 50~Mbps, thus using it beyond these limits is not advisable since no data in our possession can validate the quality of the synthetic traces.

\begin{table}[bt]
    \caption{Parameters of the proposed generative model.
    Each VR application is characterized by five parameters: two for the frame size dispersion $D_{FS}=\alpha x^\beta$, one for the 60~FPS IFI dispersion $D_{IFI}=\gamma$, two for the 30~FPS IFI dispersion $D_{IFI}=\delta x^\epsilon$.}
    \label{tab:generative_model_params}
    \begin{tabular}{l|ccccc}
    \toprule
                 & $\alpha$ & $\beta$ & $\gamma$ & $\delta$ & $\epsilon$    \\
    \midrule
    Virus Popper & 0.1784 & -0.2403 & 0.03721 & 0.01433  & 0.1764 \\
    Minecraft    & 0.1857 & -0.1872 & 0.07133 & 0.02419  & 0.2267 \\
    GE VR Tour   & 0.2554 & -0.2031 & 0.03468 & 0.01056  & 0.2756 \\
    GE VR Cities & 0.2597 & -0.2539 & 0.03457 & 0.008953 & 0.3119 \\
    \bottomrule
    \end{tabular}
\end{table}

\begin{algorithm}[tb]
    \footnotesize
    \caption{Generative model for XR traffic}
    \label{alg:model}
    \begin{algorithmic}[1]
        \REQUIRE AppName, FrameRate, DataRate
        \STATE FsAvg = DataRate / FrameRate
        \STATE IfiAvg = 1 / FrameRate
        \STATE $\alpha$, $\beta$, $\gamma$, $\delta$, $\epsilon$ = GetParameters (AppName) \COMMENT{see \cref{tab:generative_model_params}}
        \vspace{1ex}

        \STATE FsDispersion = $\alpha \cdot \mathrm{DataRate}^\beta$
        \STATE FsScale = FsDispersion $\cdot$ FsAvg
        \vspace{1ex}

        \IF{FrameRate == 60}
            \STATE IfiDispersion = $\gamma$
        \ELSIF{FrameRate == 30}
            \STATE IfiDispersion = $\delta \cdot \mathrm{DataRate}^\epsilon$
        \ELSE
            \STATE Error: only 30 and 60~FPS supported
        \ENDIF
        \STATE IfiScale = IfiDispersion $\cdot$ IfiAvg
    \end{algorithmic}
\end{algorithm}

We let different applications have separate models, obtaining a data set of 10 traces per application (half at 30~FPS, half at 60~FPS).
The parameters for all applications can be found in \cref{tab:generative_model_params} and the generative algorithm is summarized in \cref{alg:model}.

\section{Simulation Results}
\label{sec:simulation_results}

To further test the validity of the proposed model, we implemented it on top of \gls{ns3}, a popular open-source full-stack simulation software, and has been made publicly available together with the processed VR traffic traces in \texttt{CSV} format.\footnote{Available in the ns-3 app store: \url{https://apps.nsnam.org/app/bursty-app/}}
Further details on the implementation of this traffic model on \gls{ns3} can be found in~\cite{lecci21bursty}.

To test our model, we set up simulation campaigns where multiple users equipped with \glspl{hmd} communicate with a central \gls{wifi} \gls{ap}, using a wireless connection based on the IEEE~802.11ac standard.
The central \gls{ap} also acts as rendering server, generating one VR stream for each receiving \gls{sta} of the scenario.
Transmissions randomly start within the first second of simulation, avoiding that different streams start at the same time.

We show results for traffic streams imported directly from the acquired traces as well as for our model.
Since a single trace is available for each parameter combination (i.e., application, frame rate, data rate), for a fixed parameter combination, the traffic flows will all come from the same trace, although different 60~s windows are sampled to further decouple different users.
Instead, simulations running our proposed model have been repeated twice: one with the target data rate submitted to \textit{VRidge} when acquiring the corresponding trace, and one with the empirical data rate measured directly from the acquired traces (the two rates differ slightly, as can be seen from \cref{fig:measured_rate_vs_data_rate}).
This information can also be found directly into the metadata of the acquired traces, made available in \texttt{CSV} format together with our model.

\begin{table}[bt]
    \caption{List of simulation parameters}
    \label{tab:sim_params}
    \begin{tabular}{lc|lc}
    \toprule
    \textbf{Parameter} & \textbf{Value} & \textbf{Parameter}      & \textbf{Value} \\
    \midrule
    Duration           & 60 s           & RTS/CTS                 & Disabled       \\
    Distance           & 1 m            & MCS                     & VHT MCS 9      \\
    Mobility           & Fixed          & Channel Width           & 160 MHz        \\
    IP                 & v4             & Guard Interval Duration & 400 ns         \\
    Socket             & UDP            & Fragment Size           & 1472 B         \\
    \bottomrule
    \end{tabular}
\end{table}

\subsection{Model Validation}
\label{sub:model_validation}

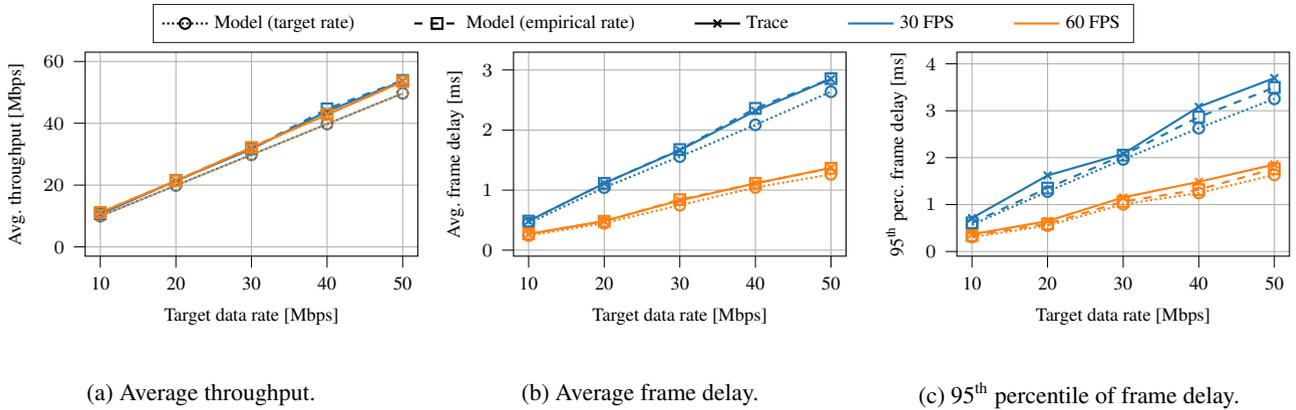
\begin{figure*}[t!]
    \setlength\fheight{0.5\columnwidth}
    \setlength\fwidth{0.7\columnwidth}
    \begin{subfigure}[t]{\textwidth}
        \centering
%
%

\definecolor{color0}{rgb}{0.12156862745098,0.466666666666667,0.705882352941177}
\definecolor{color1}{rgb}{1,0.498039215686275,0.0549019607843137}

\begin{tikzpicture}
\pgfplotsset{every tick label/.append style={font=\scriptsize}}

\begin{axis}[%
width=0,
height=0,
at={(0,0)},
scale only axis,
xmin=0,
xmax=0,
xtick={},
ymin=0,
ymax=0,
ytick={},
axis background/.style={fill=white},
legend style={legend cell align=center, align=center, draw=white!15!black, font=\scriptsize, at={(0, 0)}, anchor=center, /tikz/every even column/.append style={column sep=2em}},
legend columns=5,
]

\addplot [thick, mark=o, densely dotted, mark options={solid}, black]
table {%
0 0
};
\addlegendentry{Model (target rate)}
\addplot [thick, mark=square, dashed, mark options={solid}, black]
table {%
0 0
};
\addlegendentry{Model (empirical rate)}
\addplot [thick, mark=x, black]
table {%
0 0
};
\addlegendentry{Trace}

\addplot [thick, color0]
table {%
0 0
};
\addlegendentry{30 FPS}
\addplot [thick, color1]
table {%
0 0
};
\addlegendentry{60 FPS}

\end{axis}
\end{tikzpicture}%
    \end{subfigure}
    \\
    \hspace*{\fill}%
    \begin{subfigure}[t]{0.3\textwidth}
        \centering
\begin{tikzpicture}

\definecolor{color0}{rgb}{0.12156862745098,0.466666666666667,0.705882352941177}
\definecolor{color1}{rgb}{1,0.498039215686275,0.0549019607843137}
\definecolor{color2}{rgb}{0.172549019607843,0.627450980392157,0.172549019607843}

\begin{axis}[
  width=\fwidth,
  height=\fheight,
legend cell align={left},
legend style={
  fill opacity=0.8,
  draw opacity=1,
  text opacity=1,
  at={(0.03,0.97)},
  anchor=north west,
  draw=white!80!black
},
tick align=outside,
tick pos=left,
x grid style={white!69.0196078431373!black},
xlabel={Target data rate [Mbps]},
xmajorgrids,
xmin=-0.2, xmax=4.2,
xtick style={color=black},
xtick={0,1,2,3,4},
xticklabels={10,20,30,40,50},
y grid style={white!69.0196078431373!black},
ylabel={Avg. throughput [Mbps]},
ymajorgrids,
ymin=-3, ymax=63,
ytick style={color=black}
]


\addplot [thick, mark=o, densely dotted, mark options={solid}, color0]
table {%
0 9.94301932
1 19.8830634
2 29.82246688
3 39.7631016133333
4 49.69714072
};
\addplot [thick, mark=square, dashed, mark options={solid}, color0]
table {%
0 10.7292400933333
1 21.4770975466667
2 31.7796178666667
3 44.6672917466667
4 53.8663121333333
};
\addplot [thick, mark=x, color0]
table {%
0 10.66773776
1 21.34996736
2 31.68927312
3 43.88850576
4 53.64454576
};


\addplot [thick, mark=o, densely dotted, dash phase=1pt, color1]
table {%
0 9.93182698666667
1 19.8676968266667
2 29.8030197733333
3 39.73801888
4 49.6714852666667
};
\addplot [thick, mark=square, dashed, mark options={solid}, color1]
table {%
0 11.1514528933333
1 21.4308794266667
2 32.17264848
3 42.89704856
4 53.50695284
};
\addplot [thick, mark=x, color1]
table {%
0 11.13356416
1 21.33666744
2 32.15793704
3 42.8013236
4 53.4381848
};

\end{axis}

\end{tikzpicture}
        \caption{Average throughput.}
        \label{fig:apprate_avg_thr}
    \end{subfigure}
    \hspace*{\fill}%
    \begin{subfigure}[t]{0.3\textwidth}
        \centering
\begin{tikzpicture}

\definecolor{color0}{rgb}{0.12156862745098,0.466666666666667,0.705882352941177}
\definecolor{color1}{rgb}{1,0.498039215686275,0.0549019607843137}
\definecolor{color2}{rgb}{0.172549019607843,0.627450980392157,0.172549019607843}

\begin{axis}[
  width=\fwidth,
  height=\fheight,
legend cell align={left},
legend style={
  fill opacity=0.8,
  draw opacity=1,
  text opacity=1,
  at={(0.03,0.97)},
  anchor=north west,
  draw=white!80!black
},
tick align=outside,
tick pos=left,
x grid style={white!69.0196078431373!black},
xlabel={Target data rate [Mbps]},
xmajorgrids,
xmin=-0.2, xmax=4.2,
xtick style={color=black},
xtick={0,1,2,3,4},
xticklabels={10,20,30,40,50},
y grid style={white!69.0196078431373!black},
ylabel={Avg. frame delay [ms]},
ymajorgrids,
ymin=-0.1, ymax=3.29664578314515,
ytick style={color=black}
]


\addplot [thick, mark=o, densely dotted, mark options={solid}, color0]
table {%
0 0.448780734856489
1 1.04162812461152
2 1.55851026189737
3 2.08667661221648
4 2.63834452199996
};
\addplot [thick, mark=square, dashed, mark options={solid}, color0]
table {%
0 0.483563219215308
1 1.11261237699612
2 1.67531633677708
3 2.36147357228822
4 2.85588806654853
};
\addplot [thick, mark=x, color0]
table {%
0 0.486327528497085
1 1.11320858591382
2 1.66461427494459
3 2.31988311622688
4 2.85059850588613
};


\addplot [thick, mark=o, densely dotted, mark options={solid}, color1]
table {%
0 0.245004319829608
1 0.44760622847882
2 0.748881919313576
3 1.04393831949652
4 1.26308933541559
};
\addplot [thick, mark=square, dashed, mark options={solid}, color1]
table {%
0 0.26982158722649
1 0.480853188655837
2 0.840930970809199
3 1.11227354943921
4 1.36426346759128
};
\addplot [thick, mark=x, color1]
table {%
0 0.269639326101194
1 0.484926548721994
2 0.828020612201627
3 1.11140699589897
4 1.37099566202488
};
\end{axis}

\end{tikzpicture}
        \caption{Average frame delay.}
        \label{fig:apprate_avg_delay}
    \end{subfigure}
    \hspace*{\fill}%
    \begin{subfigure}[t]{0.3\textwidth}
        \centering
\begin{tikzpicture}

\definecolor{color0}{rgb}{0.12156862745098,0.466666666666667,0.705882352941177}
\definecolor{color1}{rgb}{1,0.498039215686275,0.0549019607843137}
\definecolor{color2}{rgb}{0.172549019607843,0.627450980392157,0.172549019607843}

\begin{axis}[
  width=\fwidth,
  height=\fheight,
legend cell align={left},
legend style={
  fill opacity=0.8,
  draw opacity=1,
  text opacity=1,
  at={(0.03,0.97)},
  anchor=north west,
  draw=white!80!black
},
tick align=outside,
tick pos=left,
x grid style={white!69.0196078431373!black},
xlabel={Target data rate [Mbps]},
xmajorgrids,
xmin=-0.2, xmax=4.2,
xtick style={color=black},
xtick={0,1,2,3,4},
xticklabels={10,20,30,40,50},
y grid style={white!69.0196078431373!black},
ylabel={95\textsuperscript{th} perc. frame delay [ms]},
ymajorgrids,
ymin=-0.1, ymax=4.24793522125123,
ytick style={color=black}
]


\addplot [thick, mark=o, densely dotted, mark options={solid}, color0]
table {%
0 0.57046617
1 1.27720394
2 1.962905595
3 2.628503555
4 3.25319277
};
\addplot [thick, mark=square, dashed, mark options={solid}, color0]
table {%
0 0.60967776
1 1.35093241
2 2.053615135
3 2.86585191
4 3.4953302
};
\addplot [thick, mark=x, color0]
table {%
0 0.71181749
1 1.62144577
2 2.08087032
3 3.08298405
4 3.69626044
};


\addplot [thick, mark=o, densely dotted, mark options={solid}, color1]
table {%
0 0.30653109
1 0.551313795
2 1.00536931
3 1.24679287
4 1.63335373
};
\addplot [thick, mark=square, dashed, mark options={solid}, color1]
table {%
0 0.337171335
1 0.58910522
2 1.065021015
3 1.31925986
4 1.765022255
};
\addplot [thick, mark=x, color1]
table {%
0 0.36535666
1 0.64794185
2 1.15232935
3 1.4849988
4 1.85375701
};
\end{axis}

\end{tikzpicture}
        \caption{95\textsuperscript{th} percentile of frame delay.}
        \label{fig:apprate_95_delay}
    \end{subfigure}
    \hspace*{\fill}%

    \caption{Simulation results for a single user streaming the \textit{Google Earth VR - Cities} application over a \gls{wifi} link.
    The statistics refer to fully received frames rather than to single fragments.}
    \label{fig:apprate}
\end{figure*}

Exhaustive simulation campaigns have been run for all four applications and five data rates at both 30~and 60~FPS, each repeated 10 times to obtain solid average statistics.
Confidence intervals are not shown as they are extremely tight.
Additional simulation parameters are shown in \cref{tab:sim_params}.

In the following section, plots will show burst-level rather than fragment-level metrics, which in case of a video stream are much more informative and bring a more realistic perspective on the quality perceived by the user.
In fact, in this case we are more interested in the performance regarding full video frames rather than single packets, and thus all packets from a burst will have to be collected before the \gls{hmd} will be able to process and show the frame to the user.

To validate our proposed model, we simulate a scenario as similar as possible to our acquisition setup, where a rendering server transmits the VR stream to a single user.
Note that the \gls{wifi} connection is able to withstand hundreds of megabits-per-second, thus a single user transmitting up to 50~Mbps is largely underutilizing the channel, allowing us to obtain unbiased results with respect to the limits of the channel capacity.
We simulated all 40 combinations of parameters (4 applications, 5 data rates, 2 frame rates), although we only show results for the 10 related to the \textit{Google Earth VR - Cities} application in \cref{fig:apprate}.

In \cref{fig:apprate_avg_thr} we show the average throughput obtained by the 3 simulation campaigns in the 10 parameter sets.
Clearly, both 30~and 60~FPS runs obtain similar results, since this metric disregards the frame rate.
In fact, both models targeting the nominal rate (shown in dots and circular markers) are perfectly superimposed on the main diagonal.
Simulations using the original traffic traces, instead, tend to have a slightly higher throughput (solid line with cross markers), as was expected by looking at \cref{fig:measured_rate_vs_data_rate}.
Since data rate, frame size, and, conversely, latency are correlated, we matched our model's data rate with the empirical one, as shown by the dashed line with square markers.
As the flexibility of our model allows us to choose an arbitrary target rate, we can see a perfect match in the computed average throughput.

In \cref{fig:apprate_avg_delay}, instead, we show the average frame delay measured from the \gls{app} layer of the \gls{ap} to the \gls{app} layer of the \gls{sta}.
Processing, encoding/decoding and other technical delays must be added to obtain the full \textit{motion-to-photon} latency, and thus the network delay should remain below 5--10~ms, as mentioned in \cref{sec:introduction}.
The most noticeable difference with respect to the previous figure is that the two frame rates are clearly separated.
This is because our reference application, described in \cref{sub:traffic_analysis}, allows us to choose a target data rate, trying to maintain a \gls{cbr} transmission during the whole duration of the stream.
This translates into frame sizes which depend directly on the frame rate, following the formula $S=R/F$ as described in \cref{sub:traffic_analysis}.
Since the channel capacity for these simulations is kept constant, doubling the frame rate halves the frame sizes which, in turn, halves the average video frame delay.
As expected, the model using the target rate slightly underestimates the average frame delay, which depends on the real application throughput, always slightly lower than the one empirically computed from the traffic traces.
Instead, similarly to the average throughput, setting the model to the trace's empirical rate yields an almost perfect match with the VR traces we acquired.
Finally, notice that the average frame delay always remains below 3~ms, far below the bound suggested by the industry experts~\cite{3gpp.26.928,itu-t-f.743.10,huaweiVrArWhitePaper,huaweiArInsight,5gAmericasServicesInnovation}.

\begin{figure*}[t!]
    \setlength\fheight{0.5\columnwidth}
    \setlength\fwidth{0.7\columnwidth}
    \begin{subfigure}[t]{\textwidth}
        \centering
%
%

\definecolor{color0}{rgb}{0.12156862745098,0.466666666666667,0.705882352941177}
\definecolor{color1}{rgb}{1,0.498039215686275,0.0549019607843137}

\begin{tikzpicture}
\pgfplotsset{every tick label/.append style={font=\scriptsize}}

\begin{axis}[%
width=0,
height=0,
at={(0,0)},
scale only axis,
xmin=0,
xmax=0,
xtick={},
ymin=0,
ymax=0,
ytick={},
axis background/.style={fill=white},
legend style={legend cell align=center, align=center, draw=white!15!black, font=\scriptsize, at={(0, 0)}, anchor=center, /tikz/every even column/.append style={column sep=2em}},
legend columns=5,
]

\addplot [thick, mark=o, densely dotted, mark options={solid}, black]
table {%
0 0
};
\addlegendentry{Model (target rate)}
\addplot [thick, mark=square, dashed, mark options={solid}, black]
table {%
0 0
};
\addlegendentry{Model (empirical rate)}
\addplot [thick, mark=x, black]
table {%
0 0
};
\addlegendentry{Trace}

\addplot [thick, color0]
table {%
0 0
};
\addlegendentry{30 FPS}
\addplot [thick, color1]
table {%
0 0
};
\addlegendentry{60 FPS}

\end{axis}
\end{tikzpicture}%
    \end{subfigure}
    \\
    \hspace*{\fill}%
    \begin{subfigure}[t]{0.3\textwidth}
        \centering
\begin{tikzpicture}

  \definecolor{color0}{rgb}{0.12156862745098,0.466666666666667,0.705882352941177}
  \definecolor{color1}{rgb}{1,0.498039215686275,0.0549019607843137}
  \definecolor{color2}{rgb}{0.172549019607843,0.627450980392157,0.172549019607843}

\begin{axis}[
  width=\fwidth,
  height=\fheight,
legend cell align={left},
legend style={
  fill opacity=0.8,
  draw opacity=1,
  text opacity=1,
  at={(0.03,0.97)},
  anchor=north west,
  draw=white!80!black
},
tick align=outside,
tick pos=left,
x grid style={white!69.0196078431373!black},
xlabel={\# Users},
xmajorgrids,
xtick={1,...,8},
xmin=0.55, xmax=8.45,
xtick style={color=black},
y grid style={white!69.0196078431373!black},
ylabel={Avg. throughput [Mbps]},
ymajorgrids,
ytick={0,100,...,500},
ymin=-10, ymax=450,
ytick style={color=black}
]


\addplot [thick, mark=o, densely dotted, mark options={solid}, color0]
table {%
1 49.69714072
2 99.0378536533333
3 148.671166266667
4 198.281240066667
5 247.89060796
6 297.483110346667
7 347.055564626667
8 396.525838733333
};
\addplot [thick, mark=square, dashed, mark options={solid}, color0]
table {%
1 53.8663121333333
2 107.341446386667
3 161.13636288
4 214.904302773333
5 268.67295016
6 322.42768396
7 376.146261
8 425.69851216
};
\addplot [thick, mark=x, color0]
table {%
1 53.64454576
2 107.00207256
3 160.5750748
4 214.41433192
5 267.981574
6 321.72018864
7 375.08760784
8 421.65836808
};


\addplot [thick, mark=o, densely dotted, dash phase=1pt, color1]
table {%
1 49.6714852666667
2 99.0183434533333
3 148.719412826667
4 198.3089586
5 248.021626106667
6 297.411775373333
7 346.9709618
8 396.52333128
};
\addplot [thick, mark=square, dashed, mark options={solid}, color1]
table {%
1 53.50695284
2 106.665824973333
3 160.20318652
4 213.62186164
5 267.172761306667
6 320.37709956
7 373.76198476
8 424.725338253333
};
\addplot [thick, mark=x, color1]
table {%
1 53.4381848
2 106.6656432
3 159.97014352
4 213.90849728
5 267.0431936
6 320.21190872
7 374.0715952
8 423.24266632
};

\end{axis}

\end{tikzpicture}
        \caption{Average throughput.}
        \label{fig:nstas_avg_thr}
    \end{subfigure}
    \hspace*{\fill}%
    \begin{subfigure}[t]{0.3\textwidth}
        \centering
\begin{tikzpicture}

    \definecolor{color0}{rgb}{0.12156862745098,0.466666666666667,0.705882352941177}
    \definecolor{color1}{rgb}{1,0.498039215686275,0.0549019607843137}
    \definecolor{color2}{rgb}{0.172549019607843,0.627450980392157,0.172549019607843}

\begin{axis}[
    width=\fwidth,
    height=\fheight,
legend cell align={left},
legend style={fill opacity=0.8, draw opacity=1, text opacity=1, draw=white!80!black},
tick align=outside,
tick pos=left,
x grid style={white!69.0196078431373!black},
xlabel={\# Users},
xmajorgrids,
xtick={1,...,8},
xmin=0.55, xmax=8.45,
xtick style={color=black},
y grid style={white!69.0196078431373!black},
ylabel={Avg. frame delay [ms]},
ymajorgrids,
ymin=-0.5, ymax=20.5,
ytick style={color=black}
]


\addplot [thick, mark=o, densely dotted, mark options={solid}, color0]
table {%
1 2.63834452199996
2 2.85229978864747
3 3.1376080350627
4 3.42498809155383
5 3.8495243804014
6 4.39356860721383
7 5.12467556142838
8 6.32635203409447
};
\addplot [thick, mark=square, dashed, mark options={solid}, color0]
table {%
1 2.85588806654853
2 3.11626717224806
3 3.43856091842508
4 3.78431557087341
5 4.29754313243498
6 4.98740817656884
7 5.96984658356712
8 10.4306220291935
};
\addplot [thick, mark=x, color0]
table {%
1 2.85059850588613
2 3.2019294324556
3 3.31382004115597
4 3.75635826298438
5 4.37572267612557
6 4.96373458163527
7 6.61753521610122
8 15.7526243435075
};


\addplot [thick, mark=o, densely dotted, mark options={solid}, color1]
table {%
1 1.26308933541559
2 1.3510488753618
3 1.46733084059819
4 1.60326567498012
5 1.77991688938167
6 1.99629218449699
7 2.30001173860572
8 2.83626906797225
};
\addplot [thick, mark=square, dashed, mark options={solid}, color1]
table {%
1 1.36426346759128
2 1.4678180794032
3 1.60468627335319
4 1.76649286758901
5 1.98546829054529
6 2.26423207842549
7 2.67339291908492
8 8.06420179048247
};
\addplot [thick, mark=x, color1]
table {%
1 1.37099566202488
2 1.46152640989279
3 1.62578574927257
4 1.7052968179483
5 1.85394651589401
6 2.12190004164415
7 2.61779987056487
8 15.0900997897286
};

\end{axis}

\end{tikzpicture}
        \caption{Average frame delay.}
        \label{fig:nstas_avg_delay}
    \end{subfigure}
    \hspace*{\fill}%
    \begin{subfigure}[t]{0.3\textwidth}
        \centering
\begin{tikzpicture}

  \definecolor{color0}{rgb}{0.12156862745098,0.466666666666667,0.705882352941177}
  \definecolor{color1}{rgb}{1,0.498039215686275,0.0549019607843137}
  \definecolor{color2}{rgb}{0.172549019607843,0.627450980392157,0.172549019607843}

\begin{axis}[
  width=\fwidth,
  height=\fheight,
legend cell align={left},
legend style={
  fill opacity=0.8,
  draw opacity=1,
  text opacity=1,
  at={(0.03,0.97)},
  anchor=north west,
  draw=white!80!black
},
tick align=outside,
tick pos=left,
x grid style={white!69.0196078431373!black},
xlabel={\# Users},
xmajorgrids,
xtick={1,...,8},
xmin=0.55, xmax=8.45,
xtick style={color=black},
y grid style={white!69.0196078431373!black},
ylabel={95\textsuperscript{th} perc. frame delay [ms]},
ymajorgrids,
ymin=-0.5, ymax=20.5,
ytick style={color=black}
]


\addplot [thick, mark=o, densely dotted, mark options={solid}, color0]
table {%
1 3.25319277
2 4.496286715
3 5.54910035
4 6.483955325
5 7.91515835
6 9.667349925
7 11.98459653
8 15.7914908
};
\addplot [thick, mark=square, dashed, mark options={solid}, color0]
table {%
1 3.4953302
2 5.021746605
3 6.13422574
4 7.23128777
5 8.91674694
6 11.01759044
7 14.09258172
8 32.421681185
};
\addplot [thick, mark=x, color0]
table {%
1 3.69626044
2 5.10511797
3 5.63143637
4 7.18466075
5 8.891815685
6 10.877762455
7 15.10806764
8 57.3459806849999
};


\addplot [thick, mark=o, densely dotted, mark options={solid}, color1]
table {%
1 1.63335373
2 2.06367666
3 2.548991765
4 3.075484765
5 3.683261655
6 4.460986825
7 5.49268275
8 7.41044284
};
\addplot [thick, mark=square, dashed, mark options={solid}, color1]
table {%
1 1.765022255
2 2.298451755
3 2.847754555
4 3.43235625
5 4.20766914
6 5.18515452
7 6.620108715
8 35.57295284
};
\addplot [thick, mark=x, color1]
table {%
1 1.85375701
2 2.18886218
3 2.87317945
4 3.12695993
5 3.57846107
6 4.487718135
7 6.0262571
8 69.98160822
};

\end{axis}

\end{tikzpicture}
        \caption{95\textsuperscript{th} percentile of frame delay.}
        \label{fig:nstas_95_delay}
    \end{subfigure}
    \hspace*{\fill}%

    \caption{Simulation results for multiple users streaming the \textit{Google Earth VR - Cities} application over a \gls{wifi} link.
    The statistics refer to fully received frames rather than to single fragments.}
    \label{fig:nstas}
\end{figure*}
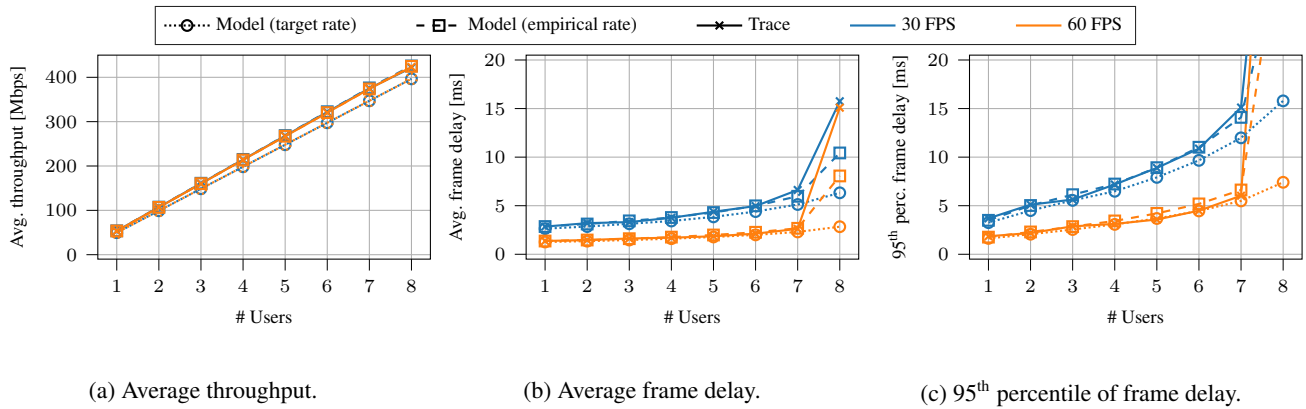

To complete this analysis, in \cref{fig:apprate_95_delay} we report the 95\textsuperscript{th} percentile delay performance of our simulations.
This metric is important as it gives an idea of the worst-case performance of the network.
In fact, only ensuring average performance is not enough to obtain a smooth and appreciable user experience, since frequent stutters in the streamed video might easily ruin the interactivity of the application and even disorient the user.
Ensuring that the 95\textsuperscript{th} percentile of the delay is within acceptable bounds allows for a more fluid and overall better experience.
In the case under analysis, it can be easily seen that both models using target and empirical rate slightly underestimate the frame delay of the acquired traces.
It is likely that the fitted Logistic distribution is not able to fully grasp the minute details of the traffic trace, making our model unable to match the real traffic.

Note that, while these results are bound to the specifications of the network under analysis (e.g., MCS, channel width, guard interval duration, fragment size, presence of RTS/CTS, \gls{wifi} standard, mobility, environment) the framework that we proposed is general.
This suggests that it can be used to study a variety of more or less complex scenarios and network architectures with different sets of parameters, assessing how they affect the end-to-end performance.

To conclude, it appears that our model is indeed able to reliably predict average statistics, while it could still be improved to better mimic slightly more advanced and specific features.
These refinements will be pursued in our future work.

\subsection{Use Case Example}
\label{sub:use_case_example}

Finally, we propose a simple example use case for our VR traffic generator.
We consider a VR arena setting, where multiple users are attached directly to a single AP streaming wirelessly.
We assume that each user requests a 50~Mbps stream and observe how many \glspl{sta} can be supported by an arena with an analogous setup.

As expected, we notice again from \cref{fig:nstas} that our model needs to be calibrated against the empirical rate of the acquired trace to yield reliable results.
In fact, from \cref{fig:nstas_avg_thr} we can see that the average throughput of the calibrated model perfectly matches the throughput of the traffic trace up to at least 8~users, where the network is able to support more than 400~Mbps.

In \cref{fig:nstas_avg_delay} it is possible to see an unstable network condition, when 8~users are trying to stream simultaneously.
It appears that the slightly higher throughput required by the trace and the empirical rate model with respect to the target rate model is enough to push the network to its limit, resulting in a sudden increase of the average frame delay, at both 30~and 60~FPS.
Focusing on the 30~FPS simulations, the plot shows that up to 6~users can be supported within the 5~ms bound, while 7~users slightly exceed this limit, and finally 8~users make the network unstable and are thus pushed over the 10~ms limit for both the trace and the model using the empirical rate.
It is important to notice that the more unstable the network, the worse the prediction accuracy of our model.
This is probably due to the simplifications that we introduced, such as the Logistic distribution and the uncorrelated samples for both the \gls{ifi} time and frame size stochastic processes.
Similarly, at 30~FPS, up to 7~users can be supported, but an additional user makes the system highly unstable and with poor prediction performance from our model.

Finally, in \cref{fig:nstas_95_delay} we show the results for the 95\textsuperscript{th} percentile of the delay.
Similarly to the average delay, this metric also shows the instability of the network for 8~users with much worse performance.
Focusing on 30~FPS, the system is able to keep the delay below the 5~ms bound only when no more than 2~users are present, whereas up to 5-6~users can be served if a 10~ms is still deemed acceptable.
Instead, at 60~FPS up to 6~users can be served while keeping the network delay within 5~ms, while the 10~ms limit is only surpassed when the network becomes unstable with 8~users.

These counterintuitive conclusions come from the fact that the application fixes a data rate, not a quality of experience.
This means that doubling the frame rate results in halving the frame size, thus reducing the perceived image quality of the streamed application, which turns into an almost halved delay.
Fixing a constant bit rate thus results in higher frame rates yielding lower latencies, at the cost of a lower image quality.

In general, there is good accordance between the results predicted by the calibrated model and the traffic traces, while the uncalibrated model often shows overly optimistic results.
When the traffic in the network increases too much and the network becomes unstable, the three simulations diverge significantly, making our synthetic traces less reliable, although this is a corner case that might be of lesser interest.

\section{XR Traffic Modeling Roadmap}
\label{sec:xr_traffic_modeling_roadmap}

Starting from the model described in the previous sections, in the following we propose an end-to-end framework to evaluate network solutions, tailored for \gls{xr} applications.
The goal is to list and detail the tasks required for the construction of such a framework, in order to encourage researchers in this field to advance with their work the state of the art, using our baseline as a valid starting point.

While~\cref{sub:first_order_stats} is devoted to highlighting our contributions, in~\cref{sub:introducing_temporal_correlation,sub:introducing_head_tracking,sub:qoe_centric_xr} we set down each additional task, describing how they can lead to the optimization of network protocols.

\subsection{Exploiting First-Order Statistics}
\label{sub:first_order_stats}

The model proposed in this paper, despite its basic functionalities, represents a solid foundation on top of which future works can iterate to develop more sophisticated strategies.
In particular, we designed an open-source, highly customizable setup (described in~\cref{sec:vr_traffic_acquisition_and_analysis}) to acquire traffic traces by sniffing the packets traveling on the local network where the experiments were conducted.

At this stage, packets are generated following first-order statistics, sampling the size and inter-frame interval from the distribution fitted on the collected data (see~\cref{sub:traffic_analysis}).
As a consequence, with this model we can emulate the creation of application frames that replicate the strategy implemented by the rendering server used in our experiments.
While this model is already useful for some applications, it lends itself to several interesting extensions, which capture other important features of the statistics of XR traffic.
As an example, in the rest of this section we discuss the importance of studying the correlation between different packets and of understanding how the movements of the user impact the generated traffic as two key areas of future improvement for our model.

\subsection{Introducing Temporal Correlation}
\label{sub:introducing_temporal_correlation}

More advanced studies can be carried out to improve the model with additional features.
One important aspect to elaborate on is the correlation among subsequent frames, or even among a specific group of frames.

As mentioned in \cref{sub:traffic_analysis}, when compressing a video stream both intra-frame and inter-frame compression techniques could be exploited, and this influences not only the structure of the packets since the type of compression greatly influences the frame size, but also the strategy to inject them into the network.
It is also possible that some manufacturers use advanced coding techniques such as Periodic Intra Refresh, as was explained in \cref{sub:traffic_analysis} for the streaming application used for our analysis, or more advanced standards such as H.265~\cite{h265} using different compression techniques.
In that case, the importance of temporal correlation might decrease, although further analysis should be carried out to ensure this.

It should be clear, by now, that the availability of a model capable of generalizing how such frame sequences are created, independently from the technical setup, is important, and the fact that each manufacturer may use its own policy represents an additional challenge.
In addition, having a model that integrates and generalizes the temporal correlation among frames would allow researchers to elaborate strategies to guarantee a certain level of latency and throughput, for example by giving different priorities and scheduling options to different types of packets.

For applications with constant delay requirements and high values of \gls{fps}, a solution could be to buffer (at the device side or at the rendering server) specific packets associated with keyframes, in order to improve the encoding process.
This would require stable network performance and an application capable of communicating directly with the network, e.g., exploiting cross-layer solutions, to be aware of any change of the link quality that would trigger specific countermeasures or improvements, if applicable.

\subsection{Introducing Head Tracking}
\label{sub:introducing_head_tracking}

A further improvement of the model should exploit the information on movement tracking, in particular related to the head, for all \gls{6dof}.
In this case, sniffing the packets traveling through the network might not be enough, and we thus need to gather information from different sensors (e.g., gyroscope, accelerometer and compass), that could be integrated into the device used to interact with the virtual world.

With respect to VRidge, the software that we used to make our phone acting as a VR headset and our PC as a rendering server, the developers provide an API for this purpose.\footnote{https://github.com/RiftCat/vridge-api}
By connecting to the head tracking endpoint, the software provides positional, rotational, or combined data, and even the possibility of modifying phone tracking data in real time before it is used for the rendering step.

This is important because, by aligning the motion trace with the traffic generated by the application, it can be determined whether there is a correlation between a certain movement of the user and the corresponding drop in the reception of packets, or other network-related events.
For example, knowing the direction of the physical movement of the user might help mmWave wireless systems (such as 802.11ad/ay) keep beam alignment between the \gls{ap} and the user device, thus limiting the risk of abrupt connection interruptions if the line of sight is lost.

It has to be highlighted that this approach could benefit every communication infrastructures that can be used to deliver XR content, as user tracking data can be exploited at different layers of the protocol stack.

\subsection{Full Traffic Emulator}
\label{sub:full_traffic_emulator}

The last step to further increase the fidelity (but also the complexity) of the traffic model is to fully characterize and emulate all the different information sub-flows and how they interact with each other.
For example, as shown in \cref{fig:stem} and explained in \cref{sub:traffic_analysis}, the VR stream comprises both \gls{dl} and \gls{ul} messages containing information such as video frames, head tracking information, and feedback.

A full-blown emulator would send all this information to and from the user, reacting accordingly whenever a packet is lost or corrupted, or when communication delays are present.
This level of detail requires a much more in-depth analysis of the transmission protocol of a real \gls{xr} application, understanding all the consequences of erratic and unexpected behaviors of the network.

Such a precise model would be extremely useful when running large simulation campaigns as it would give the most accurate and reliable results.
However, the amount of work required to analyze and reproduce a realistic behavior would be extremely high.

\subsection{QoE-centric XR}
\label{sub:qoe_centric_xr}

As highlighted thoroughly in the previous paragraphs, the final goal of all these approaches is to guarantee high-level performance to the final user.
In particular, in the \gls{xr} domain, we tend to measure the performance in terms of overall satisfaction of the customers, referred to as \gls{qoe}, and, to the best of our knowledge, there is no standardized way to evaluate these metrics.

In our case, besides the quality of the shown image, also the latency of the communication between the \gls{hmd} and the rendering server can make a difference (especially if the latter is in the cloud), considering that cybersickness has a huge impact on the user experience.
For this reason, researchers should be encouraged to design algorithms that guarantee stable and constant performance, taking into account that the traffic in the network varies depending on the application and user activity.

Moreover, since in a common scenario we have different users, there may be a need to support different traffic categories at the same time in the same network.
This requires a system able to fairly distribute resources among the flows, where learning algorithms could be implemented to orchestrate every operation, either from a network or from an application perspective.

Given a certain condition of the user, or other available information, the algorithm could predict the \gls{qoe} trend and act accordingly in case of an anticipated performance drop.
At this point on the roadmap, the network design should focus on the user, trying to guarantee a stable experience also when \gls{vbr} flows are considered.
In fact, in a \gls{cbr} flow (much easier to handle from a network point of view), the perceived image quality can be affected in case of a scene with a large amount of action and details.
In this case, it may be difficult to fit everything at a fixed rate and, as a consequence, the user experiences a downgrade in terms of quality.
This further highlights the need for novel solutions, able to tackle these problems by trading off system complexity and \gls{qoe}.

\section{Conclusions}
\label{sec:conclusions}


In this paper we described the current state of the art regarding the telecommunication aspects needed to support high-quality \gls{xr} streaming, mainly focusing on the challenges needed to obtain faithful traffic models that the community could use to test protocols and optimize networks.

We then proceeded to acquire over 4~hours of VR traffic, study in detail this type of traffic, and propose a model to generate synthetic traffic traces, while also making freely available to the community both our implementation and the VR dataset.

Finally, we show some results on the predictive power of our model, while also acknowledging its weak points.
Furthermore, we provided an example use case where multiple users coexist in the same network, naively sharing radio resources up to its collapse.
Further work could better study effective scheduling strategies for \gls{xr} traffic streams, possibly coexisting with other applications in the same network while also ensuring robustness in case of fluctuating channel quality.
Also, the model could be tested and validated for higher values of FPS, by collecting and analyzing additional traces at 90~FPS or higher.

With this work, we hope to pave the way for the research community to start working towards the optimization and support of this specific type of traffic, given the extreme interest from the main standard bodies and the most prominent telecommunication industries.


\section*{Acknowledgment}
We want to thank the \textit{RiftCat} team for patiently answering all the disclosable technical questions we asked, allowing us to improve our work.

\bibliographystyle{IEEEtran}
\bibliography{bibl}

\begin{IEEEbiography}[{\includegraphics[width=1in,height=1.25in,clip,keepaspectratio]{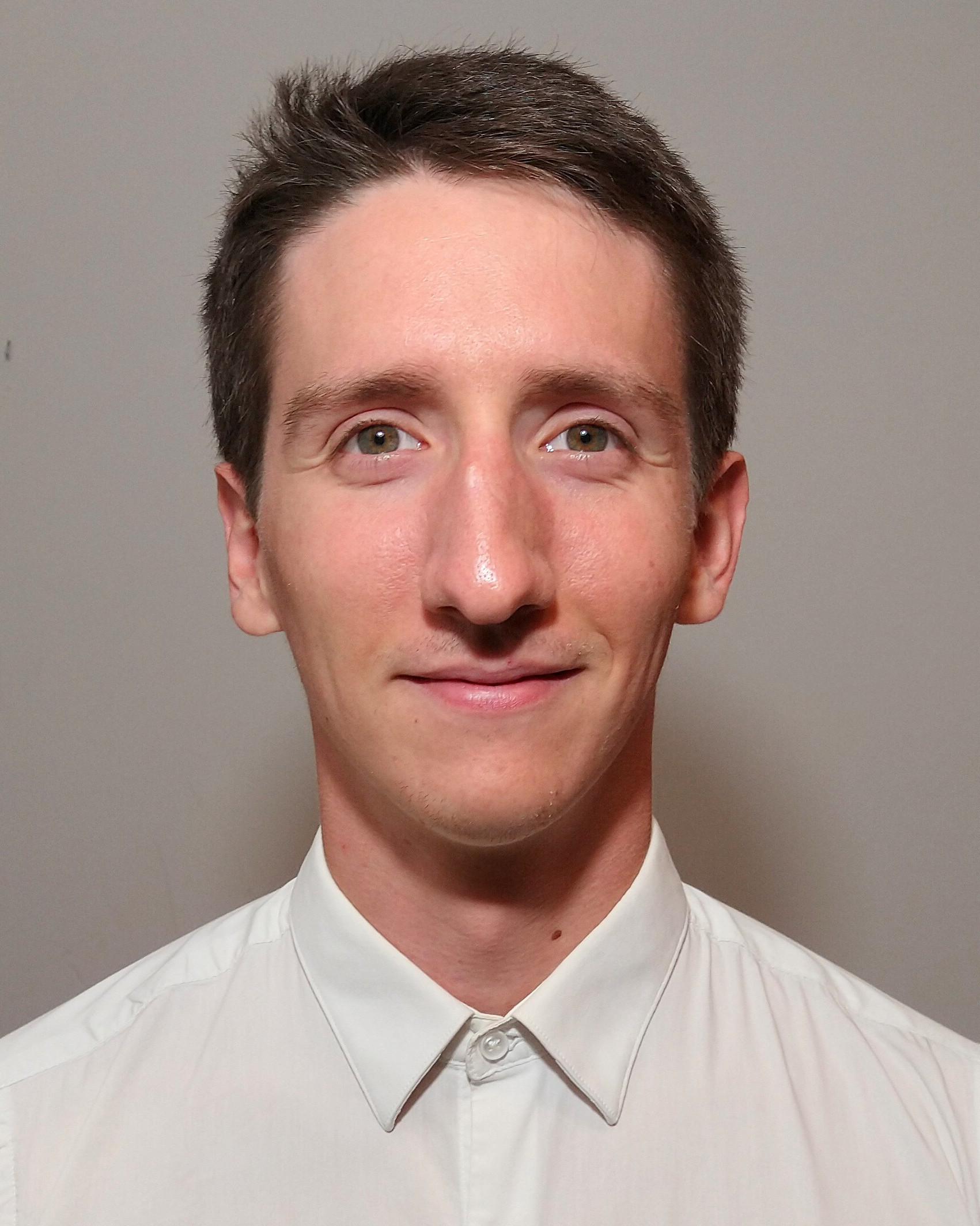}}]{Mattia Lecci}
    (Graduate Student Member, IEEE) received the B.Sc. degree (Hons.) in information engineering and the M.Sc. degree (Hons.) in telecommunication engineering from the University of Padova, Italy, in 2016 and 2018, respectively, where he is currently pursuing the Ph.D. degree in information engineering.

    He was a Guest Researcher with the National Institute for Standards and Technology (NIST) in 2018. His main research activities are channel modeling for the mmWave frequency band, MAC scheduling for WiGig technologies, applied machine learning for communications, virtual reality traffic modeling, and open-source software development.
\end{IEEEbiography}
\begin{IEEEbiography}[{\includegraphics[width=1in,height=1.25in,clip,keepaspectratio]{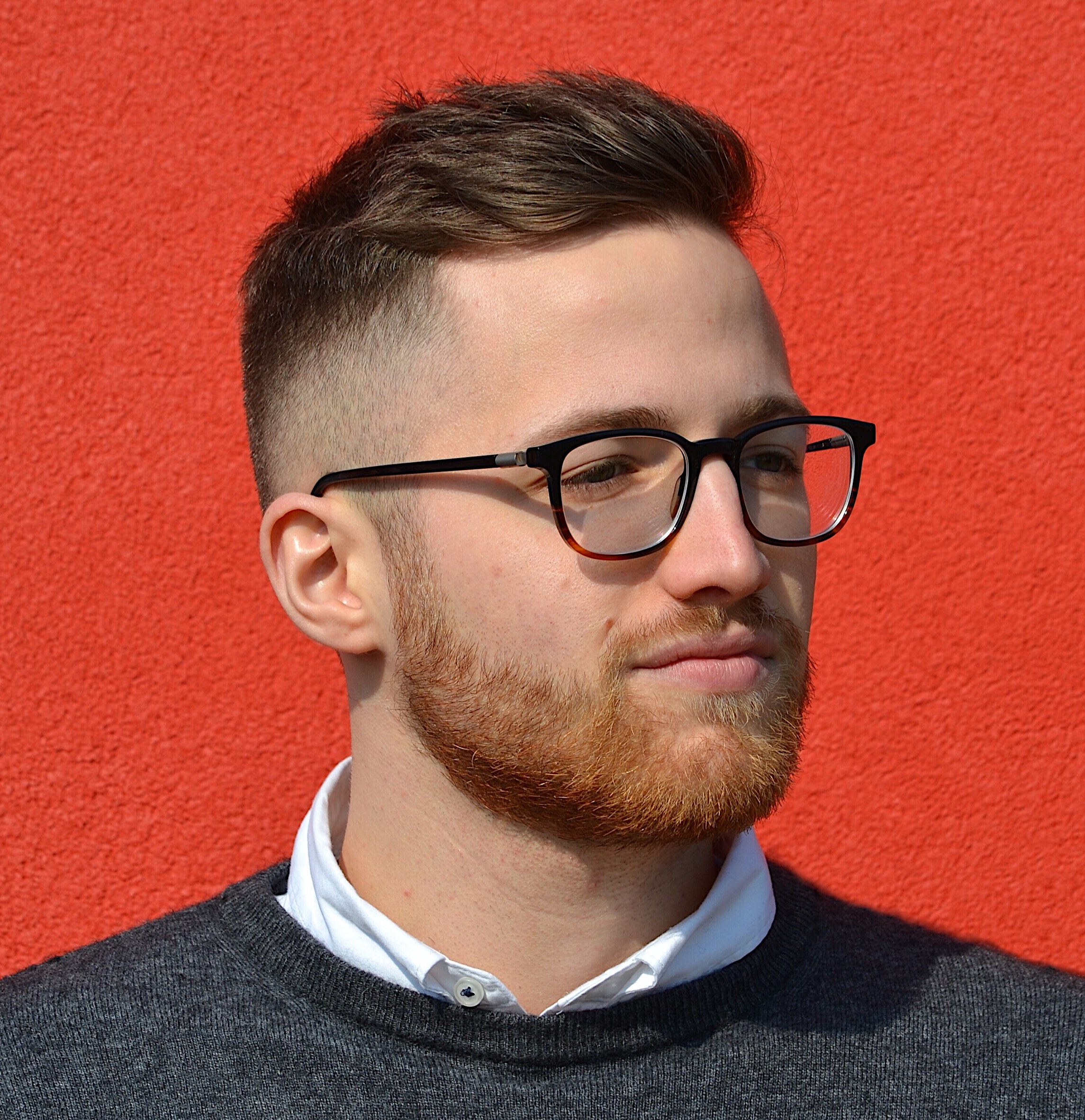}}]{Matteo Drago}
    (Graduate Student Member, IEEE) received his B.Sc. (2016) and M.Sc. (2019) in Telecommunication
    Engineering from the University of Padova, Italy. Since October 2019, he has been a Ph.D. Student at the University of Padova. He visited Nokia Bell Labs, Dublin, in 2018, working on QoS provisioning in 60 GHz networks. His research interests are in the study of the next generation of vehicular networks operating at millimeter-wave.
\end{IEEEbiography}
\begin{IEEEbiography}[{\includegraphics[width=1in,height=1.25in,clip,keepaspectratio]{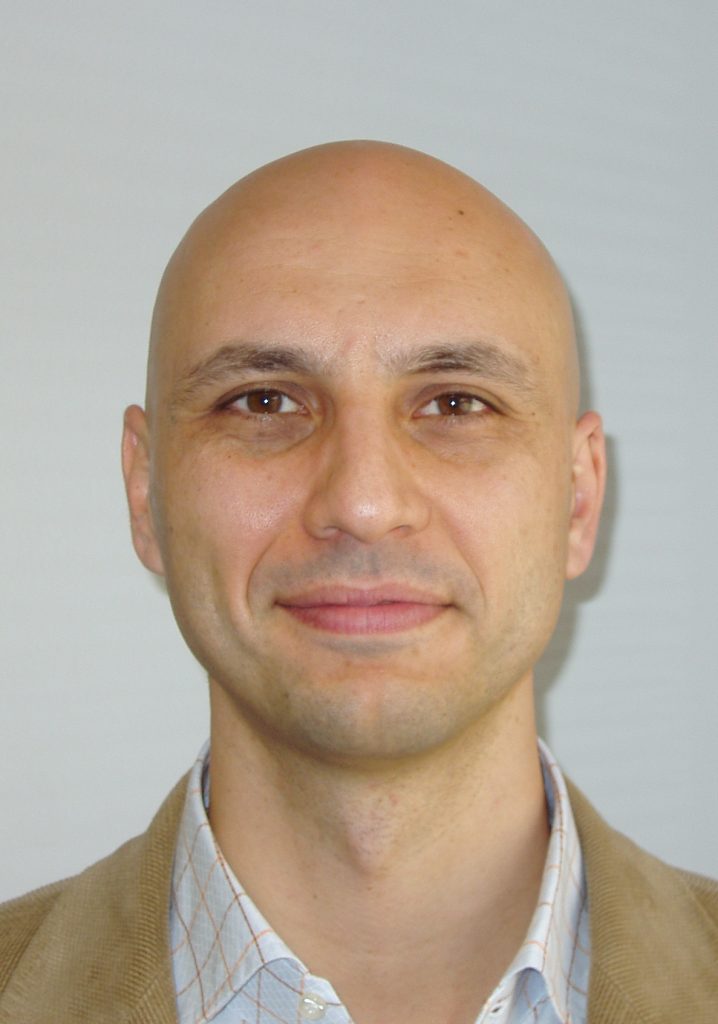}}]{Andrea Zanella}(Senior Member, IEEE) received the Laurea degree in computer engineering from the University of Padova, Italy, in 1998, and the Ph.D. degree in 2001. In 2000, he spent nine months with Prof. Mario Gerla's research team at the University of California, Los Angeles (UCLA). He is currently a Full Professor with the Department of Information Engineering (DEI), University of Padova. He is one of the coordinators of the SIGnals and NETworking (SIGNET) research lab. His long-established research activities are in the fields of protocol design, optimization, and performance evaluation of wired and wireless networks. He has been serving as a Technical Area Editor for the IEEE INTERNET OF THINGS JOURNAL and an Associate Editor for the IEEE TRANSACTIONS ON COGNITIVE COMMUNICATIONS AND NETWORKING, the IEEE COMMUNICATIONS SURVEYS AND TUTORIALS, and Digital Communications and Networks
\end{IEEEbiography}
\begin{IEEEbiography}[{\includegraphics[width=1in,height=1.25in,clip,keepaspectratio]{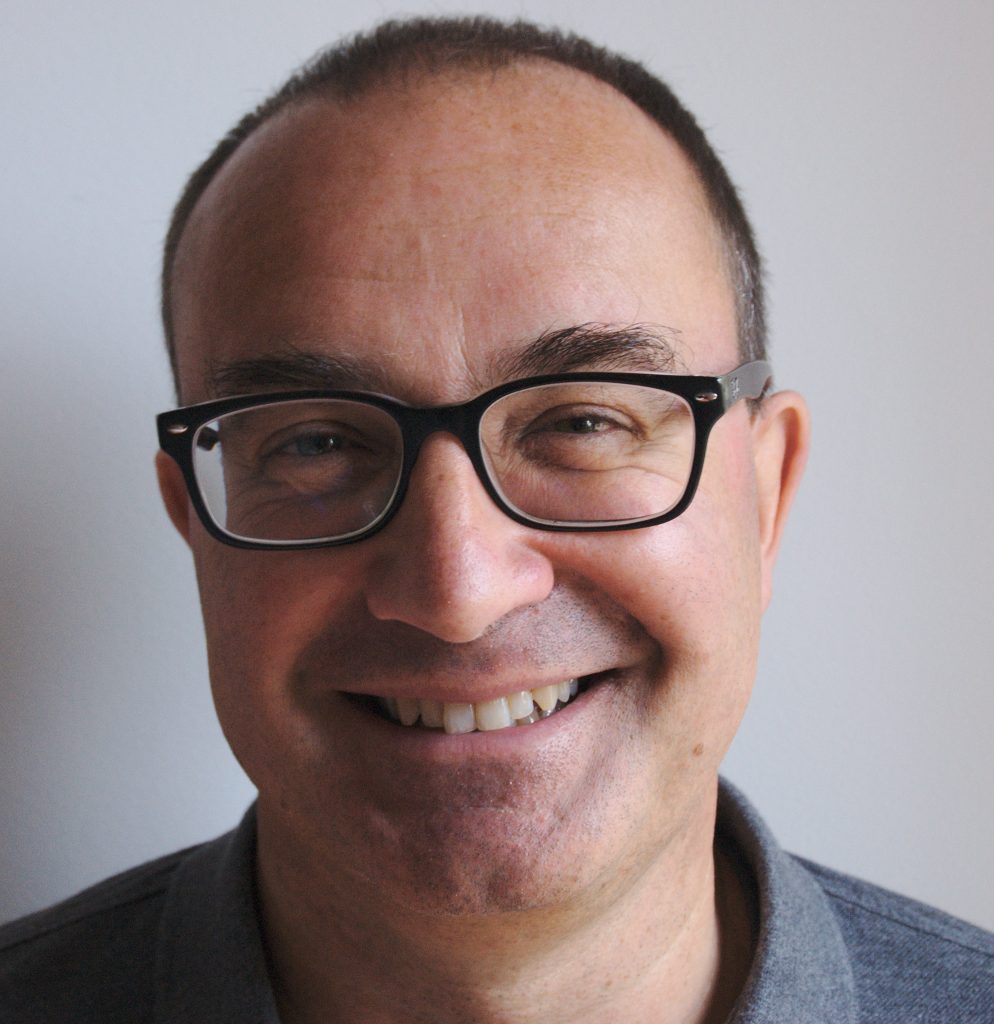}}]{Michele Zorzi}
    (Fellow, IEEE) received the Laurea and Ph.D. degrees in electrical engineering from the University of Padova, Italy, in 1990 and 1994, respectively.

    From 1992 to 1993, he was on leave at the University of California at San Diego (UCSD). In 1993, he joined the Dipartimento di Elettronica e Informazione, Politecnico di Milano, Italy. After spending three years with the Center for Wireless Communications, UCSD, in 1998 he joined the School of Engineering, University of Ferrara, Italy, where he became a Professor in 2000. Since November 2003, he has been a Faculty Member with the Department of Information Engineering, University of Padova. His current research interests include performance evaluation in mobile communications systems, the Internet of Things, cognitive communications and networking, 5G mmWave cellular systems, vehicular networks, and underwater communications and networks.

    Dr. Zorzi received several awards from the IEEE Communications Society, including the Best Tutorial Paper Award in 2008 and 2019, the Education Award in 2016, the Stephen O. Rice Best Paper Award in 2018, and the Joseph LoCicero Award for Exemplary Service to Publications in 2020. He was the Editor-in-Chief of the IEEE Wireless Communications Magazine from 2003 to 2005, the IEEE TRANSACTIONS ON COMMUNICATIONS from 2008 to 2011, and the IEEE TRANSACTIONS ON COGNITIVE COMMUNICATIONS AND NETWORKING from 2014 to 2018. He has served the IEEE Communications Society as a Member-at-Large of the Board of Governors from 2009 to 2011 and from 2021 to 2023, as the Director of Education from 2014 to 2015, and as the Director of Journals from 2020 to 2021.
\end{IEEEbiography}

\EOD

\end{document}